\documentclass[12pt]{article}
\usepackage{graphicx}
\usepackage{lscape}
\usepackage{epsfig}
\usepackage{citesort}
\usepackage{amssymb}
\usepackage{amsmath}
\usepackage{multirow}

\newcommand{\be}{\begin{equation}}
\newcommand{\ee}{\end{equation}}
\newcommand{\bea}{\begin{eqnarray}}
\newcommand{\eea}{\end{eqnarray}}
\newcommand{\nn}{\nonumber}

\def\R1{\varepsilon_1}
\def\E8{\varepsilon_8}

\def\ga{\gamma}

\def\lb{\Lambda_b}

\def\s1{\hat s}
\def\ds{\displaystyle}

\newcommand{\bd}{\begin{displaymath}}
\newcommand{\ed}{\end{displaymath}}

\newcommand{\f}{\frac}

\def\R1{\varepsilon_1}
\def\E8{\varepsilon_8}

\def\ga{\gamma}

\def\ds{\displaystyle}
\def\beq{\begin{equation}}
\def\eeq{\end{equation}}
\def\bea{\begin{eqnarray}}
\def\eea{\end{eqnarray}}
\def\beeq{\begin{eqnarray}}
\def\eeeq{\end{eqnarray}}

\def\vel{\left|}
\def\ver{\right|}
\def\nnb{\nonumber}
\def\ga{\left(}
\def\dr{\right)}

\def\rar{\rightarrow}
\def\nnb{\nonumber}

\def\ba{\begin{array}}
\def\ea{\end{array}}

\def\xis0{{\Xi^{*0}}}

\def\g5{\gamma_5}

\def\es{\!\!\! &=& \!\!\!}

\def\ar{&+& \!\!\!}
\def\ek{&-& \!\!\!}

\def\cp{&\times& \!\!\!}

\newcommand{\al}{\alpha_s}

\begin{document}
\title{
         {\Large
                 {\bf Constraint on  compactification scale via recently observed baryonic $\Lambda _b
\rightarrow \Lambda \ell^+ \ell^-$ channel and analysis of the $\Sigma _b \rightarrow \Sigma  \ell^+ \ell^-$
transition in SM and UED scenario
                 }
         }
      }
\author{\vspace{1cm}\\
{\small  K. Azizi$^1$ \thanks {e-mail: kazizi@dogus.edu.tr}\,\,, S. Kartal$^2$ \thanks
{e-mail: sehban@istanbul.edu.tr}\,\,, N. Kat{\i}rc{\i}$^1$ \thanks
{e-mail: nkatirci@dogus.edu.tr}\,\,, A. T. Olgun$^2$ \thanks
{e-mail: a.t.olgun@gmail.com.tr}\,\,, Z. Tavuko\u glu$^2$ \thanks{
e-mail: z.tavukoglu@gmail.com.tr}}  \\
{\small $^1$ Department of Physics, Do\u gu\c s University,
Ac{\i}badem-Kad{\i}k\"oy, 34722 \.{I}stanbul, Turkey}\\
{\small $^2$ Department of Physics, \.{I}stanbul University,
Vezneciler, 34134 \.{I}stanbul, Turkey}\\
}
\date{}

\begin{titlepage}
\maketitle
\thispagestyle{empty}
\begin{abstract}
We obtain a lower limit on the compactification scale of extra dimension  via comparison of the branching ratio in the baryonic  $\Lambda _b
\rightarrow \Lambda \mu^+ \mu^-$ decay channel recently measured by CDF collaboration and our previous theoretical study. We also use the newly available form factors calculated via 
light cone QCD sum rules in full
theory to analyze the flavour changing neutral current process of the $\Sigma_b \rightarrow \Sigma
\ell^+ \ell^-$ in universal extra dimension scenario in the presence of a single extra compact dimension.  We calculate various physical quantities like branching ratio, forward-backward asymmetry, baryon polarizations
 and double lepton polarization asymmetries defining the decay channel under consideration. We also compare the obtained predictions  with those of the standard model. 
\end{abstract}
~~~PACS number(s): 12.60-i,  13.30.-a, 13.30.Ce, 14.20.Mr
\end{titlepage}


\section{Introduction}
The CDF Collaboration at Fermilab has recently reported the first observation of the baryonic flavour changing neutral current (FCNC) decay $\Lambda _b\rightarrow \Lambda \mu^+ \mu^-$ with $24$ 
signal events and a statistical significance of $5.8~\sigma$ \cite{CDF}. This event as the first FCNC observation in baryonic sector has stimulated both experimental and theoretical studied in this area. The LHCb collaboration
at CERN has also started to study this decay channel \cite{LHCb}.  Comparison of the theoretical and phenomenological predictions on related physical observables with experimental data can help us get valuable
information not only about the internal structure  of the participating particles, strong interaction and other parameters of the standard model (SM) but about the new physics effects. Such comparison  leads to
put constraints on the parameters existing in many new physics scenarios beyond the SM (BSM).

The FCNC transitions are very important frameworks to indirectly search for extra dimensions and Kaluza Klein (KK) particles as new physics effects. In the past, putting constraints on the compactification scale,
 $1/R$ of extra dimensions and mass of KK modes was passable only via comparison of the experimental data on physical observables with theoretical predictions in mesonic sector. By the above mentioned developments, now,
it is possible to get knowledge on these parameters also in FCNC baryonic decay channels. Our first task in the present study is to put constraint on the compactification scale of extra dimension 
by comparing the experimental data on the branching fraction of the $\Lambda _b\rightarrow \Lambda \mu^+ \mu^-$ and our theoretical prediction \cite{azizi-katirci} in universal extra dimension (UED) framework  with a single
 extra dimension called Applequist-Cheng-Dobrescu (ACD) model (For more information about the model and idea of extra dimension (ED) see \cite{ACD,Appelquist,Antoniadis,Antoniadis2,Arkani,Arkani1,Randall1,Randall2}).
 Note that this decay channel was studied in detail in  SM in \cite{aliev-kazem}.

In the second and main part of the present study, we work out the other baryonic FCNC  $\Sigma_b \rightarrow \Sigma \ell^+ \ell^- $ transition in the context of UED may will be in agenda of experiments in future.
 We use the form factors, very recently calculated via light cone QCD sum rules in full theory \cite{sigma}, as the main ingredients in this channel.  The order of branching ratio on this channel reported in \cite{sigma} shows that
this decay channel is also accessible at LHC. We use the transition form factors enrolled to the low energy effective Hamiltonian to calculate many physical observables related to the decay channel under 
consideration. Particularly, we evaluate the branching ratio, forward-backward asymmetry, baryon polarizations and double lepton polarization asymmetries both in the SM and UED and compare our results on
the considered physical quantities  obtained via these two models. The UED model has also been applied to many channels mainly in mesonic sector
(see for instance \cite{Buras,KK,Bashiry,nihan,Bayar,KK2,Yu-Ming,Aliev-Savci,De-Fazio,Sirvanli-azizi,Pak,Aliev-Sirvanli,Aslam,Colangelo,Giri} and references therein).

The layout of the article is as follows. In next section, we find a lower limit on the compactification scale via comparing the experimental result on the branching ratio of the  
 $\Lambda_b \rightarrow \Lambda\mu^+ \mu^-$ and theoretical prediction. In section 3, we evaluate the $\Sigma_b \rightarrow \Sigma  \ell^+ \ell^-$ transition in UED model and calculate the corresponding physical 
quantities. In this section, we also  numerically  analyze the observables defining the transition under consideration and compare the obtained results with SM predictions.
Last section encompasses our discussions and conclusions.

\section{ Constraint on the Compactification Factor via $\Lambda_b \rightarrow \Lambda \mu^+ \mu^-$ Decay Channel} 
In ED models \cite{Antoniadis,Antoniadis2,Arkani,Arkani1,Randall1,Randall2}, gravity can travel in the higher dimensional bulk. This give rise to KK towers of massive spin-2 graviton excitations
or KK gravitons whose possible destination can be a tour along a circle of radius $R$ called  size of the extra dimension  and return to where they began. The mass difference between subsequent KK particles is
of order $1/R$. In UED model, the SM fields (both gauge bosons and fermions) are also allowed to propagate in the extra dimensions \cite{ACD,Appelquist}. As a result of interactions among the SM and KK particles,
the Wilson coefficients entering effective Hamiltonian become functions of compactification scale $1/R$ (we will come back to this point in next section). Hence, it will be of great importance to put constraint on
this factor.

The lower bound of compactification factor has been put mainly comparing the experimental data with theoretical calculations in mesonic channel, electroweak precision tests and some cosmological constraints.
Analysis of the $B\rightarrow X_s\gamma$ decay channel  and  anomalous magnetic moment depict  that when $1/R\geq 300~GeV$, the experimental data are in  good agreements with the UED model predictions \cite{ikiuc}. 
In \cite{ACD,Appelquist}, based on also  the electroweak precision tests, it has been  found  that the lower limit for compactification scale is $%
250~GeV$  when $M_{Higgs}\geq250~GeV$ denoting larger KK contributions to the low energy FCNC transitions,
 and $300~GeV$  when $M_{Higgs}\leq250~GeV$. According to  \cite{Gogoladze:2006br} and \cite{Cembranos:2006gt}, again the 
electroweak precision measurements as well as
some cosmological constraints give rise to  $500~GeV$ for the lower limit on compactification scale. Contributing the leading order
(LO) and next-to-next-to-leading order (NNLO) corrections due to the exchange of KK modes  also to the ${\cal B}(B \rightarrow X_s \gamma)$ transition in \cite{Haisch:2007vb} has lead to $600~GeV$ as lower bound on 
$1/R$. Moreover, the ATLAS collaboration at CERN  has set a $600~GeV$ on the lower bound of $1/R$,
for values of the compression scale between $2$ and $40$, implying $730~ GeV$ for lower bound of  the mass of the KK gluons \cite{ATLAS}.
However, very recently, the authors of   \cite{colangeloR} have found that the theoretical result on ${\cal B}(B \rightarrow K \eta \gamma)$ matches with
experimental data if $1/R\gtrsim 250~GeV$ as far as they  consider a single UED. This is lower than the bound provided by other processes \cite{rlimit}.
 But when they add the second dimension (with 2 UEDs), they find $\simeq400~GeV$ for the lower limit of the 
compactification factor. 

 As we previously mentioned, now we have  the first experimental measurement on the branching ratio of $\Lambda_b \rightarrow \Lambda \mu^+\mu^-$, i.e.,
 $B(\Lambda_b \rightarrow \Lambda \mu^+\mu^-)=[1.73 \pm 0.42$(stat)$ \pm 0.55$(syst)$] \times 10^{-6}$
\cite{CDF}. This gives a possibility to obtain a lower bound on the compactification scale $1/R$ in baryonic sector by comparison between this experimental result and our previous theoretical prediction 
\cite{azizi-katirci} but only when a single UED is considered. The process $B \rightarrow K \eta \gamma$ is described by only one Wilson coefficient $C_7^{eff}$ whose explicit expression is available in UED model
with 2 EDs. However, in our case the Effective Hamiltonian describing the $\Lambda_b \rightarrow \Lambda \mu^+\mu^-$ channel contains additional coefficients, $C_9^{eff}$ and  $C_{10}$
 (for details see next section) whose values have not been known
in UED with 2 EDs yet. Hence, it is now possible to find a lower limit on $1/R$ via baryonic FCNC $\Lambda_b \rightarrow \Lambda \mu^+\mu^-$ process  in UED model with a single ED. 
The comparison is made in Figure 1
where we have considered the errors of form factors and uncertainties of other input parameters in theoretical calculations.

\begin{figure}[h!]
\begin{center}
\includegraphics{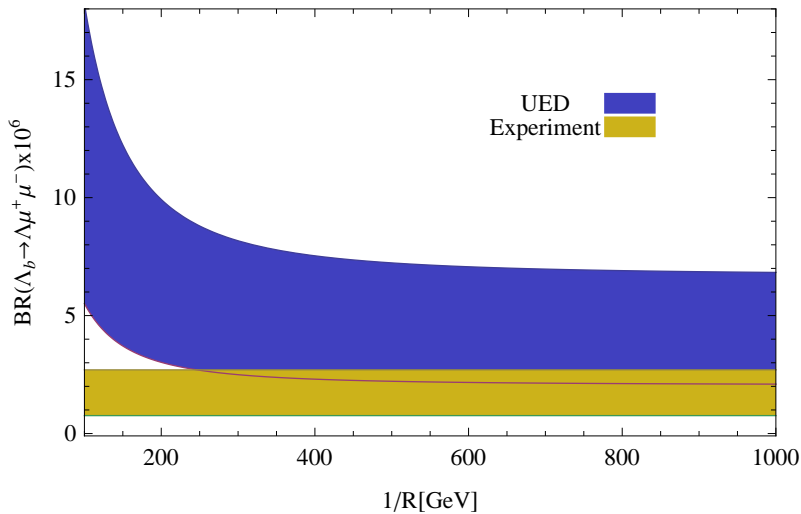}
\end{center}
\caption{ Comparison between the experimental result \cite{CDF} and theoretical prediction \cite{azizi-katirci} on the branching ratio of  $\Lambda_b \rightarrow \Lambda \mu^+ \mu^-$ channel. \label{fig1}}
\end{figure}

From this figure, we obtain an approximately $250~GeV$ for the lower bound of $1/R$  which is in a good consistency with the result of \cite{colangeloR} when only one UED is taken into account. To improve our result, one should take the
 effects
of second ED in the process under consideration and this will be possible when the explicit form of additional Wilson coefficients $C_9^{eff}$ and  $C_{10}$ are known.

\section{ The $\Sigma_{b}\rightarrow \Sigma \ell^{+}\ell^{-}$  Transition in UED }

\subsection{The Effective Hamiltonian and Transition Matrix Elements}

The FCNC transition of the $\Sigma_b \rightarrow \Sigma  \ell^+\ell^-$ proceeds via loop-level $b \rar s l^+ l^- $ transition whose effective Hamiltonian can be written as
 \bea \label{Heff} {\cal H}^{eff} &=& {G_F \alpha_{em} V_{tb}
V_{ts}^\ast \over 2\sqrt{2} \pi} \Bigg[ C_9^{eff}
\bar{s}\gamma_\mu (1-\gamma_5) b \, \bar{\ell} \gamma^\mu \ell +
C_{10}  \bar{s} \gamma_\mu (1-\gamma_5) b \, \bar{\ell}
\gamma^\mu
\gamma_5 \ell \nnb \\
&-&  2 m_b C_7^{eff}  {1\over q^2} \bar{s} i \sigma_{\mu\nu}
(1+\gamma_5) b \, \bar{\ell} \gamma^\mu \ell \Bigg]~, \eea
where $G_F$ is the Fermi weak  coupling constant, $V_{ij}$ are the
Cabibbo-Kobayashi-Maskawa (CKM) matrix elements, $\alpha_{em}$ is
the fine structure constant;  and
$C_7^{eff}$, $C_9^{eff}$ and  $C_{10}$   are  Wilson coefficients. The transition  amplitude of hadronic decay channel under consideration is defined as
 \bea\label{genlik}
{\cal M}=\langle \Sigma(p) \mid{\cal H}^{eff}\mid \Sigma_b(p+q)\rangle.
\eea
As a result of this procedure, we get the following transition  matrix elements parameterized in terms of transition form factors:
\bea\label{transmatrix1} \langle
\Sigma(p) \mid  \bar s \gamma_\mu (1-\gamma_5) b \mid \Sigma_b(p+q)\rangle\es
\bar {u}_\Sigma(p) \Big[\gamma_{\mu}f_{1}(q^{2})+{i}
\sigma_{\mu\nu}q^{\nu}f_{2}(q^{2}) + q^{\mu}f_{3}(q^{2}) \nnb \\
\ek \gamma_{\mu}\gamma_5
g_{1}(q^{2})-{i}\sigma_{\mu\nu}\gamma_5q^{\nu}g_{2}(q^{2})
- q^{\mu}\gamma_5 g_{3}(q^{2})
\vphantom{\int_0^{x_2}}\Big] u_{\Sigma_{b}}(p+q)~,\nnb \\
\eea
and,
\bea\label{transmatrix2}
\langle \Sigma(p)\mid \bar s i \sigma_{\mu\nu}q^{\nu} (1+ \gamma_5)
b \mid \Sigma_b(p+q)\rangle \es\bar{u}_\Sigma(p)
\Big[\gamma_{\mu}f_{1}^{T}(q^{2})+{i}\sigma_{\mu\nu}q^{\nu}f_{2}^{T}(q^{2})+
q^{\mu}f_{3}^{T}(q^{2}) \nnb \\
\ar \gamma_{\mu}\gamma_5
g_{1}^{T}(q^{2})+{i}\sigma_{\mu\nu}\gamma_5q^{\nu}g_{2}^{T}(q^{2})
+ q^{\mu}\gamma_5 g_{3}^{T}(q^{2})
\vphantom{\int_0^{x_2}}\Big] u_{\Sigma_{b}}(p+q)~,\nnb \\
\eea where $f^{(T)}_i$ and $g^{(T)}_i$ ($i$ runs from $1$ to $3$) are form factors; and  $u_{\Sigma_{b}}$ and $u_{\Sigma}$ are  spinors of $\Sigma_b$
 and $\Sigma$ baryons, respectively. These form factors as the main inputs in analysis of the $\Sigma_{b}\rightarrow \Sigma \ell^{+}\ell^{-}$ have been very recently calculated in full QCD via light
 cone QCD sum rules in \cite{sigma}. By full QCD, we mean full theory of QCD  without any approximation like heavy quark effective theory (HQET) limit. The fit function of transition form factors is given as \cite{sigma}:
\begin{eqnarray}
f^{(T)}_i(q^2)[g^{(T)}_i(q^2)]=\frac{a}{(1-\frac{q^2}{m_{fit}^2})}+\frac{b}{(1-\frac{q^2}{m_{fit}^2})^2},
\label{parametrization1}
\end{eqnarray}
where the fit parameters $a$, $b$, and $m_{fit}$ are
presented in Table~\ref{tab:13}.
\begin{table}[h]
\renewcommand{\arraystretch}{1.5}
\addtolength{\arraycolsep}{3pt}
$$
\begin{array}{|c|c|c|c|c|c|}

\hline \hline
                & \mbox{a} & \mbox{b}  & m_{fit}& q^2=0
                \\
\hline
 f_1            &  -0.035 &   0.13   &  5.1   &  0.095 \pm 0.017  \\
 f_2            &  0.026  &  -0.081  &  5.2   & -0.055 \pm 0.012  \\
 f_3            &   0.013 &  -0.065  &  5.3   & -0.052 \pm 0.016  \\
 g_1            &  -0.031 &   0.15   &  5.3   &  0.12 \pm 0.03    \\
 g_2            &  0.015  &  -0.040  &  5.3   & -0.025 \pm 0.008  \\
 g_3            &  0.012  &  -0.047  &  5.4   & -0.035 \pm 0.009  \\
 f_1^{T}        &  1.0    &   -1.0   &  5.4   &  0.0\pm0.0        \\
 f_2^{T}        &  -0.29  &   0.42   &  5.4   &  0.13 \pm 0.04    \\
 f_3^{T}        &  -0.24  &   0.41   &  5.4   &  0.17 \pm 0.05    \\
 g_1^{T}        &  0.45   &   -0.46  &  5.4   & -0.010\pm 0.003   \\
 g_2^{T}        &  0.031  &   0.055  &  5.4   &  0.086 \pm 0.024  \\
 g_3^{T}        &  -0.011 &  -0.18   &  5.4   & -0.19 \pm 0.06    \\
\hline \hline
\end{array}
$$
\caption{Parameters appearing in  the fit function of the  form
factors, $f_{1}$, $f_{2}$, $f_{3}$, $g_{1}$, $g_{2}$, $g_{3}$,
$f^T_{1}$, $f^T_{2}$, $f^T_{3}$, $g^T_{1}$, $g^T_{2}$ and $g^T_{3}$
in full theory for $\Sigma_{b}\rightarrow \Sigma\ell^{+}\ell^{-}$ together with the
  values of the form factors at $q^2=0$ \cite{sigma}.} \label{tab:13}
\renewcommand{\arraystretch}{1}
\addtolength{\arraycolsep}{-1.0pt}
\end{table}

After the above comments about the amplitude and transition matrix elements, we go on to discuss the source of main differences  between UED and SM models. Such differences belong to the Wilson coefficients entered the 
effective Hamiltonian. As we previously mentioned the KK particles in UED models interact with themselves as well as the SM particles in the bulk, giving rise to modifications in the SM versions
of the Wilson coefficients although the form of effective Hamiltonian remain unchanged.  Each Wilson coefficient in UED scenario is
defined in terms of a SM part $F_0(x_t)$ and  extra periodic functions  $F_n(x_{t},x_n)$ coming from new interactions, i.e., 
\bea F(x_t,1/R)=F_0(x_t)+\sum_{n=1}^{\infty}F_n(x_t,x_n).\label{function} \eea
Here, 
$x_{t}=\frac{m_{t}^{2}}{M_{W}^{2}}$,   $x_n=\displaystyle{m_n^2 \over m_W^2}$, and  $m_n=\displaystyle{n \over R} $.  Also, $m_t$,  $M_{W}$ and $m_n$ are  masses of the top
quark,   $W$ boson and KK particles (non-zero modes), respectively. The  Wilson coefficients $C_7^{eff}$, $C_9^{eff}$ and  $C_{10}$ have been calculated in UED  in the presence of a single
ED and SM models in \cite{Buras,KK,C7eff,Misiak,Munz}. The $C_9^{eff}$ which is a function of $\hat{s}'=\frac{q^2}{m_b^2}$ with $4 m_l^2\leq q^2\leq (m_{\Sigma_b}-m_{\Sigma})^2$ and compactification scale, is given as
 \bea \label{wilson-C9eff}
C_9^{eff}(\hat{s}',1/R) & = & C_9^{NDR}(1/R)\eta(\hat s') + h(z, \hat s')\left( 3
C_1 + C_2 + 3 C_3 + C_4 + 3
C_5 + C_6 \right) \nonumber \\
& & - \f{1}{2} h(1, \hat s') \left( 4 C_3 + 4 C_4 + 3
C_5 + C_6 \right) \nonumber \\
& & - \f{1}{2} h(0, \hat s') \left( C_3 + 3 C_4 \right)
+ \f{2}{9} \left( 3 C_3 + C_4 + 3 C_5 +
C_6 \right), \eea
where \bea \eta(\hat s') & = & 1 + \f{\al(\mu_b)}{\pi}\,
\omega(\hat s'), \eea with
 \bea \label{omega-shatp}
\omega(\hat s') & = & - \f{2}{9} \pi^2 - \f{4}{3}\mbox{Li}_2(\hat
s') - \f{2}{3}
\ln \hat s' \ln(1-\hat s') - \f{5+4\hat s'}{3(1+2\hat s')}\ln(1-\hat s') - \nonumber \\
& &  \f{2 \hat s' (1+\hat s') (1-2\hat s')}{3(1-\hat s')^2
(1+2\hat s')} \ln \hat s' + \f{5+9\hat s'-6\hat s'^2}{6 (1-\hat
s') (1+2\hat s')}, \eea
and
\bea
\alpha_s(x)=\frac{\alpha_s(m_Z)}{1-\beta_0\frac{\alpha_s(m_Z)}{2\pi}\ln(\frac{m_Z}{x})}.\eea
Here, $\alpha_s(m_Z)=0.118$ and $\beta_0=\frac{23}{3}$.
At $\mu_b$ scale we have
 \bea \label{CJ} C_j=\sum_{i=1}^8 k_{ji}
\eta^{a_i} \qquad (j=1,...6) \vspace{0.2cm}, \eea
where 
 \bea \eta \es
\frac{\alpha_s(\mu_W)} {\alpha_s(\mu_b)}~,\eea
\be\frac{}{}
   \label{coefficients}
\begin{array}{rrrrrrrrrl}
a_i = (\!\! & \f{14}{23}, & \f{16}{23}, & \f{6}{23}, & -
\f{12}{23}, &
0.4086, & -0.4230, & -0.8994, & 0.1456 & \!\!)  \vspace{0.1cm}, 
\end{array}
\ee
and  \be\frac{}{}
   \label{KJI}
\begin{array}{rrrrrrrrrl}
k_{1i} = (\!\! & 0, & 0, & \f{1}{2}, & - \f{1}{2}, &
0, & 0, & 0, & 0 & \!\!),  \vspace{0.1cm} \\
k_{2i} = (\!\! & 0, & 0, & \f{1}{2}, &  \f{1}{2}, &
0, & 0, & 0, & 0 & \!\!),  \vspace{0.1cm} \\
k_{3i} = (\!\! & 0, & 0, & - \f{1}{14}, &  \f{1}{6}, &
0.0510, & - 0.1403, & - 0.0113, & 0.0054 & \!\!),  \vspace{0.1cm} \\
k_{4i} = (\!\! & 0, & 0, & - \f{1}{14}, &  - \f{1}{6}, &
0.0984, & 0.1214, & 0.0156, & 0.0026 & \!\!),  \vspace{0.1cm} \\
k_{5i} = (\!\! & 0, & 0, & 0, &  0, &
- 0.0397, & 0.0117, & - 0.0025, & 0.0304 & \!\!) , \vspace{0.1cm} \\
k_{6i} = (\!\! & 0, & 0, & 0, &  0, &
0.0335, & 0.0239, & - 0.0462, & -0.0112 & \!\!).  \vspace{0.1cm} \\
\end{array}
\ee

The function, $h(y,\hat s')$ is given as
 \bea \label{h-phasespace} h(y,
\hat s') & = & -\f{8}{9}\ln\f{m_b}{\mu_b} - \f{8}{9}\ln y +
\f{8}{27} + \f{4}{9} x \\
& & - \f{2}{9} (2+x) |1-x|^{1/2} \left\{
\begin{array}{ll}
\left( \ln\left| \f{\sqrt{1-x} + 1}{\sqrt{1-x} - 1}\right| - i\pi
\right), &
\mbox{for } x \equiv \f{4z^2}{\hat s'} < 1 \nonumber \\
2 \arctan \f{1}{\sqrt{x-1}}, & \mbox{for } x \equiv \f {4z^2}{\hat
s'} > 1,
\end{array}
\right. \\
\eea where $y=1$ or $y=z=\frac{m_c}{m_b}$ and, \bea h(0, \hat s')
& = & \f{8}{27} -\f{8}{9} \ln\f{m_b}{\mu_b} - \f{4}{9} \ln \hat s'
+ \f{4}{9} i\pi.\eea
The $C_9^{NDR}(1/R)$ in \eqref{wilson-C9eff} is
expressed as
 \bea \label{C9NDR}C_9^{NDR}(1/R) & = & P_0^{NDR} +
\f{Y(x_t,1/R)}{\sin^2\theta_W} -4 Z(x_t,1/R) + P_E E(x_t,1/R), \eea where
 $P_0^{NDR}=2.60 \pm 0.25$,
$\sin^2\theta_W=0.23$ \cite{Misiak,Munz} and NDR is the
abbreviation, used for naive dimensional regularization. Due to  smallness of  the $P_E$, the last term in  \eqref{C9NDR} is neglected and remaining functions,  $Y(x_t,1/R)$ and $Z(x_t,1/R)$
 are defined in the following way:
 \bea \label{Y-function}
Y(x_t,1/R) \es Y_0(x_t)+\sum_{n=1}^\infty C_n(x_t,x_n)~,\eea where
\bea \label{Y0-function} Y_0(x_t) \es \frac{x_t}{8} \left[
\frac{x_t -4}{x_t -1}+\frac{3 x_t}{(x_t-1)^2} \ln x_t \right]~,
\eea and
 \bea
\label{CN} \sum_{n=1}^\infty C_n(x_t,x_n) = \frac{x_t(7-x_t)}{ 16
(x_t-1)} - \frac{\pi m_W R x_t}{16
(x_t-1)^2} \left[3(1+x_t)J(R,-1/2)+(x_t-7)J(R,1/2) \right]~.\nnb\\
\eea with \bea \label{JR-function} J(R,\alpha)=\int_0^1 dy \,
y^\alpha \left[ \coth (\pi m_W R \sqrt{y})-x_t^{1+\alpha}
\coth(\pi m_t R \sqrt{y}) \right]~. \eea
The $Z(x_t,1/R)$ is defined as
 \bea \label{Z-function}
 Z(x_t,1/R) \es Z_0(x_t)+\sum_{n=1}^\infty
C_n(x_t,x_n)~, \eea where
 \bea Z_0(x_t) \es \frac{18 x_t^4-163 x_t^3+259
x_t^2 -108 x_t}{144 (x_t-1)^3} \nnb +\left[\frac{32 x_t^4-38
x_t^3-15 x_t^2+18 x_t}{72
(x_t-1)^4} - \frac{1}{9}\right] \ln x_t .\\ \nnb \\
\eea

The Wilson coefficient, $C_{10}$ can be written as
 \bea \label{wilson-C10} C_{10}(1/R)= - \frac{Y(x_t,1/R)}{\sin^2 \theta_W}~. \eea

Finally, in leading log approximation, the Wilson coefficient
$C_7^{eff}(1/R)$ is given as 
  \bea
\label{wilson-C7eff} C_7^{eff}(\mu_b, 1/R) \es
\eta^{\frac{16}{23}} C_7(\mu_W, 1/R)+ \frac{8}{3} \left(
\eta^{\frac{14}{23}} -\eta^{\frac{16}{23}} \right) C_8(\mu_W,
1/R)+C_2 (\mu_W) \sum_{i=1}^8 h_i \eta^{a_i}~, \nnb\\ \eea
where
\bea
 C_2(\mu_W)=1~,~~ C_7(\mu_W, 1/R)=-\frac{1}{2}
D^\prime(x_t,1/R)~,~~ C_8(\mu_W, 1/R)=-\frac{1}{2}
E^\prime(x_t,1/R)~ . \eea The functions,  $D^\prime (x_t,1/R)$ and
$E^\prime (x_t,1/R)$ are given as: \bea D^\prime
(x_t,1/R)=D^\prime_0(x_t)+\sum_{n=1}^{\infty}D^\prime_n(x_t,x_n),~~~~~E^\prime
(x_t,1/R)=E^\prime_0(x_t)+\sum_{n=1}^{\infty}E^\prime_n(x_t,x_n)~,
\eea where  $D^\prime_0(x_t)$ and $E^\prime_0(x_t)$ 
have expresions \bea \label{Dprime0} D^\prime_0(x_t) \es -
\frac{(8 x_t^3+5 x_t^2-7 x_t)}{12 (1-x_t)^3}
+ \frac{x_t^2(2-3 x_t)}{2(1-x_t)^4}\ln x_t~, \\ \nnb \\
\label{Eprime0} E^\prime_0(x_t) \es - \frac{x_t(x_t^2-5 x_t-2)}
{4(1-x_t)^3} + \frac{3 x_t^2}{2 (1-x_t)^4}\ln x_t~, \eea and the
functions representing KK contributions are,
 \bea
\label{DN-prime} \sum_{n=1}^{\infty}D^\prime_n(x_t,x_n) \es
\frac{x_t[37 - x_t(44+17 x_t)]}{72 (x_t-1)^3} \nnb \\
\ar \frac{\pi m_W R}{12} \Bigg[ \int_0^1 dy \, (2 y^{1/2}+7
y^{3/2}+3 y^{5/2}) \, \coth (\pi m_WR \sqrt{y}) \nnb \\
\ek \frac{x_t (2-3 x_t) (1+3 x_t)}{(x_t-1)^4}J(R,-1/2)\nnb \\
\ek \frac{1}{(x_t-1)^4} \{ x_t(1+3 x_t)+(2-3 x_t)
[1-(10-x_t)x_t] \} J(R, 1/2)\nnb \\
\ek \frac{1}{(x_t-1)^4} [ (2-3 x_t)(3+x_t) + 1 - (10-x_t) x_t]
J(R, 3/2)\nnb \\
\ek \frac{(3+x_t)}{(x_t-1)^4} J(R,5/2) \Bigg]~, \nnb\\
\eea \bea \label{EN-prime}
\sum_{n=1}^{\infty}E^\prime_n(x_t,x_n)\es
\frac{x_t[17+(8-x_t)x_t]}
{24 (x_t-1)^3} \nnb \\
\ar \frac{\pi m_W R}{4} \Bigg[\int_0^1 dy \, (y^{1/2}+
2 y^{3/2}-3 y^{5/2}) \, \coth (\pi m_WR \sqrt{y}) \nnb \\
\ek {x_t(1+3 x_t) \over (x_t-1)^4}J(R,-1/2)\nnb \\
\ar \frac{1}{(x_t-1)^4} [ x_t(1+3 x_t) - 1 + (10-x_t)x_t] J(R, 1/2)\nnb \\
\ek \frac{1}{(x_t-1)^4} [(3+x_t)-1+(10-x_t)x_t) ]J(R, 3/2)\nnb \\
\ar{(3+x_t) \over  (x_t-1)^4} J(R,5/2)\Bigg]~. \eea
The coefficients  $h_i$ in Eq.\eqref{wilson-C7eff} are
given by the following values \cite{Misiak,Munz}:
   \be\frac{}{}
   \label{coefficients}
\begin{array}{rrrrrrrrrl}
h_i = (\!\! & 2.2996, & - 1.0880, & - \f{3}{7}, & - \f{1}{14}, &
-0.6494, & -0.0380, & -0.0186, & -0.0057 & \!\!).
\end{array}
\ee

\subsection{Branching Ratio}

Having the decay amplitude in Eq.\eqref{genlik}, the $1/R$-dependent double differential decay rate  is obtained as \cite{Aliev-Savci,BR,BR1}:
 \bea \frac{d^2\Gamma}{d\hat
sdz}(z,\hat s,1/R) = \frac{G_F^2\alpha^2_{em} m_{\Sigma_b}}{16384
\pi^5}| V_{tb}V_{ts}^*|^2 v \sqrt{\lambda} \, \Bigg[{\cal
T}_0(\hat s,1/R)+{\cal T}_1(\hat s,1/R) z +{\cal T}_2(\hat s,1/R)
z^2\Bigg]~, \nnb\\ \label{dif-decay}
\eea
where $\hat s=\frac{q^2}{ m_{\Sigma_b}^2}$,  $z=\cos\theta$ and $\theta$ is the angle between
momenta of lepton $l^+$ and $\Sigma_b$ in the center of mass of leptons. Here,
$\lambda=\lambda(1,r,\hat s)=(1-r-\hat s)^2-4r\hat s$ is the usual triangle
function, $v=\sqrt{1-\frac{4 m_\ell^2}{q^2}}$ is the lepton velocity and $r= m^2_{\Sigma}/m^2_{\Sigma_b}$.
The ${\cal T}_i(\hat s,1/R)$ functions are given as:
 \bea {\cal T}_0(\hat s,1/R) \es 32 m_\ell^2
m_{\Sigma_b}^4 \hat s (1+r-\hat s) \ga \vel D_3 \ver^2 +
\vel E_3 \ver^2 \dr \nnb \\
\ar 64 m_\ell^2 m_{\Sigma_b}^3 (1-r-\hat s) \, \mbox{\rm Re} [D_1^\ast
E_3 + D_3
E_1^\ast] \nnb \\
\ar 64 m_{\Sigma_b}^2 \sqrt{r} (6 m_\ell^2 - m_{\Sigma_b}^2 \hat s)
{\rm Re} [D_1^\ast E_1] \nnb \\
\ar 64 m_\ell^2 m_{\Sigma_b}^3 \sqrt{r} \Big( 2 m_{\Sigma_b} \hat s
{\rm Re} [D_3^\ast E_3] + (1 - r + \hat s)
{\rm Re} [D_1^\ast D_3 + E_1^\ast E_3]\Big) \nnb \\
\ar 32 m_{\Sigma_b}^2 (2 m_\ell^2 + m_{\Sigma_b}^2 \hat s) \Big\{ (1
- r + \hat s) m_{\Sigma_b} \sqrt{r} \,
\mbox{\rm Re} [A_1^\ast A_2 + B_1^\ast B_2] \nnb \\
\ek m_{\Sigma_b} (1 - r - \hat s) \, \mbox{\rm Re} [A_1^\ast B_2 +
A_2^\ast B_1] - 2 \sqrt{r} \Big( \mbox{\rm Re} [A_1^\ast B_1] +
m_{\Sigma_b}^2 \hat s \,
\mbox{\rm Re} [A_2^\ast B_2] \Big) \Big\} \nnb \\
\ar 8 m_{\Sigma_b}^2 \Big\{ 4 m_\ell^2 (1 + r - \hat s) +
m_{\Sigma_b}^2 \Big[(1-r)^2 - \hat s^2 \Big]
\Big\} \ga \vel A_1 \ver^2 +  \vel B_1 \ver^2 \dr \nnb \\
\ar 8 m_{\Sigma_b}^4 \Big\{ 4 m_\ell^2 \Big[ \lambda + (1 + r -
\hat s) \hat s \Big] + m_{\Sigma_b}^2 \hat s \Big[(1-r)^2 - \hat s^2 \Big]
\Big\} \ga \vel A_2 \ver^2 +  \vel B_2 \ver^2 \dr \nnb \\
\ek 8 m_{\Sigma_b}^2 \Big\{ 4 m_\ell^2 (1 + r - \hat s) -
m_{\Sigma_b}^2 \Big[(1-r)^2 - \hat s^2 \Big]
\Big\} \ga \vel D_1 \ver^2 +  \vel E_1 \ver^2 \dr \nnb \\
\ar 8 m_{\Sigma_b}^5 \hat s v^2 \Big\{ - 8 m_{\Sigma_b} \hat s \sqrt{r}\,
\mbox{\rm Re} [D_2^\ast E_2] +
4 (1 - r + \hat s) \sqrt{r} \, \mbox{\rm Re}[D_1^\ast D_2+E_1^\ast E_2]\nnb \\
\ek 4 (1 - r - \hat s) \, \mbox{\rm Re}[D_1^\ast E_2+D_2^\ast E_1] +
m_{\Sigma_b} \Big[(1-r)^2 -\hat s^2\Big] \ga \vel D_2 \ver^2 + \vel
E_2 \ver^2\dr \Big\},\nnb \\
\eea
\bea {\cal T}_1(\hat s,1/R) &=& -16
m_{\lb}^4\s1 v \sqrt{\lambda}
\Big\{ 2 Re(A_1^* D_1)-2Re(B_1^* E_1)\nn\\
&+& 2m_{\lb}
Re(B_1^* D_2-B_2^* D_1+A_2^* E_1-A_1^*E_2)\Big\}\nn\\
&+&32 m_{\lb}^5 \s1~ v \sqrt{\lambda} \Big\{
m_{\lb} (1-r)Re(A_2^* D_2 -B_2^* E_2)\nn\\
&+& \sqrt{r} Re(A_2^* D_1+A_1^* D_2-B_2^*E_1-B_1^* E_2)\Big\}\;,
\eea
 \bea {\cal T}_2(\hat s,1/R)\es - 8 m_{\Sigma_b}^4 v^2 \lambda \ga
\vel A_1 \ver^2 + \vel B_1 \ver^2 + \vel D_1 \ver^2
+ \vel E_1 \ver^2 \dr \nnb \\
\ar 8 m_{\Sigma_b}^6 \hat s v^2 \lambda \Big( \vel A_2 \ver^2 + \vel
B_2 \ver^2 + \vel D_2 \ver^2 + \vel E_2 \ver^2  \Big)~, \eea where,
 \bea \label{coef-decay-rate} A_1 \es \frac{1}{q^2}\ga
f_1^T+g_1^T \dr \ga -2 m_b C_7^{eff}(1/R)\dr + \ga f_1-g_1 \dr C_9^{eff}(\hat s,1/R) \nnb \\
A_2 \es A_1 \ga 1 \rar 2 \dr ~,\nnb \\
A_3 \es A_1 \ga 1 \rar 3 \dr ~,\nnb \\
B_1 \es A_1 \ga g_1 \rar - g_1;~g_1^T \rar - g_1^T \dr ~,\nnb \\
B_2 \es B_1 \ga 1 \rar 2 \dr ~,\nnb \\
B_3 \es B_1 \ga 1 \rar 3 \dr ~,\nnb \\
D_1 \es \ga f_1-g_1 \dr C_{10}(1/R) ~,\nnb \\
D_2 \es D_1 \ga 1 \rar 2 \dr ~, \nnb\\
D_3 \es D_1 \ga 1 \rar 3 \dr ~,\nnb \\
E_1 \es D_1 \ga g_1 \rar - g_1 \dr ~,\nnb \\
E_2 \es E_1 \ga 1 \rar 2 \dr ~,\nnb \\
E_3 \es E_1 \ga 1 \rar 3 \dr ~,
 \eea

Performing integral over $z$ in Eq.\eqref{dif-decay} in the interval $[-1,1]$,  the $1/R$-dependent differential decay rate  with respect to only $\hat s$ is obtained
as follows:
 \bea
\frac{d\Gamma}{d \hat s} (\hat s,1/R)= \frac{G_F^2\alpha^2_{em} m_{\Sigma_b}}{8192
\pi^5}| V_{tb}V_{ts}^*|^2 v \sqrt{\lambda} \, \Bigg[{{\cal T}_0(\hat s,1/R)
+\frac{1}{3} {\cal T}_2(\hat s,1/R)}\Bigg]~. \label{decayrate} \eea
To obtain the $1/R$-dependent branching ratio, we need to perform integral over $\hat s$ in the above equation in the interval,
 $\frac{4 m_\ell^2}{m_{\Sigma_b}^2}\le\hat s \le (1-\sqrt{r})^2$ and  multiply the obtained result by the lifetime of the $\Sigma_b$ baryon.
 As the lifetime of $\Sigma_b$ baryon has not exactly known, we  take it the same as the lifetime of b baryon admixture
($\Lambda_b$, $\Xi_b$, $\Sigma_b$, $\Omega_b$). To numerically analyze the branching ratio, we need also some inputs, whose   values are taken as $m_t=167~GeV$, $m_W=80.4~GeV$, $m_Z=91~GeV$,
$m_b=4.8~GeV$,
 $m_c=1.46~GeV$, $\mu_b=5~GeV$, $\mu_W=80.4~GeV$,
$| V_{tb}V_{ts}^\ast|=0.041$, $G_F = 1.17 \times 10^{-5}~ GeV^{-2}$, $\alpha_{em}=\frac{1}{137}$,
 $\tau_{\Sigma_b}=1.391\times 10^{-12}~s$, $m_\Sigma =
1.192~GeV$, $m_{\Sigma_{b}} = 5.807~ GeV$,  $m_\mu = 0.1056~GeV$ and $m_\tau = 1.776~GeV$ \cite{PDG}. 

The branching ratio of decay channel under consideration on $1/R$ is plotted in Figure 2 for both SM and UED models as well as for two lepton channels. As the results of $e$ are  close to those of
$\mu$ channel, we do not present the results in $e$ channel.
\begin{figure}[h!]
\label{fig1}
\centering
\begin{tabular}{ccc}
\epsfig{file=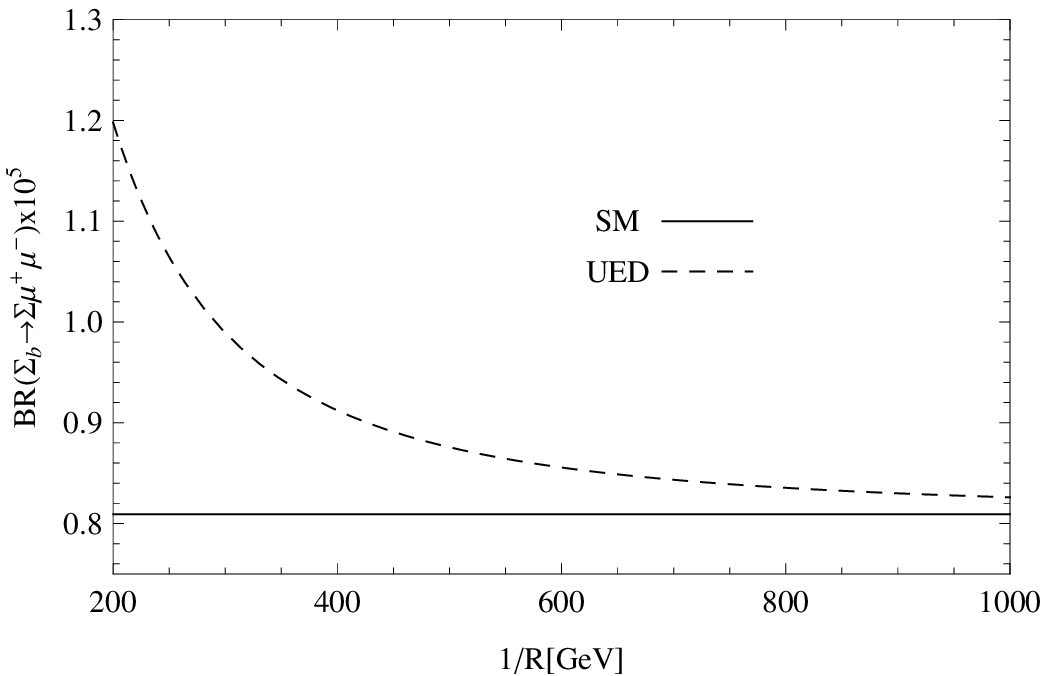,width=0.45\linewidth,clip=} &
\epsfig{file=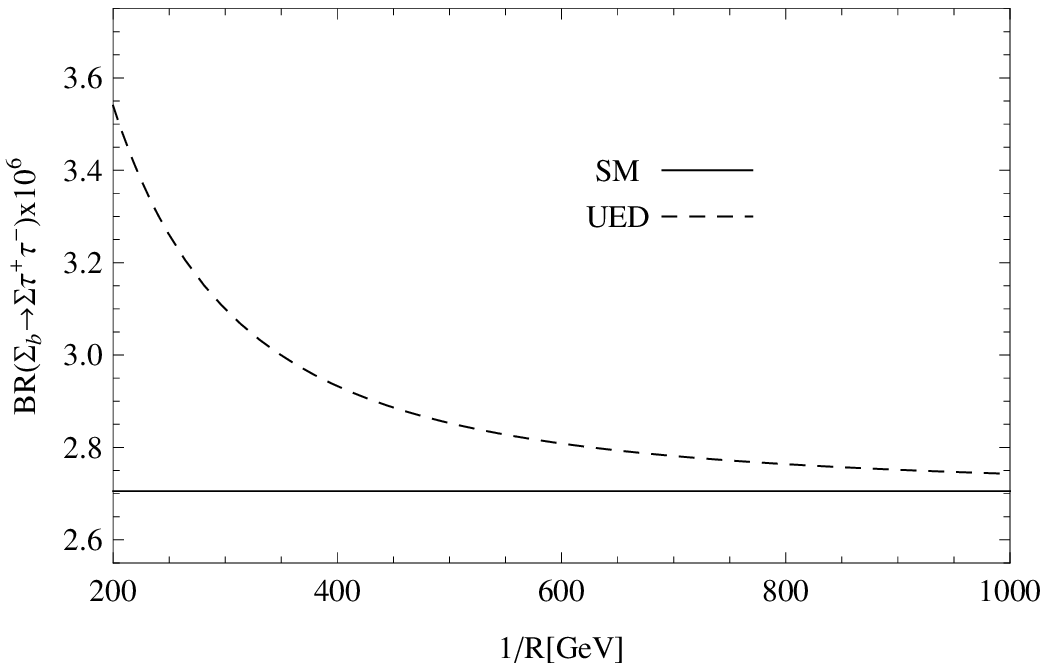,width=0.45\linewidth,clip=}
\end{tabular}
\caption{The $1/R$ dependence of the branching ratio for $\Sigma_{b}\rightarrow \Sigma \ell^{+}\ell^{-}$.}
\end{figure}
From Figure 2, we see that
\begin{itemize}
 \item there are sizable difference between the UED and SM predictions in small values of $1/R$ in both lepton channel. Such discrepancies can be considered
as indications of existing KK excitations. In higher values of compactification factor the UED results approaches to those of the SM and
two models have approximately the same predictions. 
\item The value of branching ratio at every point in $\mu$ channel is bigger than that of the $\tau$. This is an expected result.
\item The order of branching ratios show that this decay channel is accessible at LHC.
\end{itemize}

\subsection{Lepton Forward Backward Asymmetry}
The lepton forward-backward asymmetry (${\cal A}_{FB}$) which is one of useful tools to search for new physics effects is  defined as;
\bea {\cal A}_{FB} = \frac{N_f-N_b}{N_f+N_b}.
 \eea
Here, $N_f$ symbolizes the number of moving particles to forward direction, while $N_b$
represents the number of moving particles to backward direction. In technique language, the above formula leads to 
\bea {\cal A}_{FB} (\hat s,1/R)=
\frac{\ds{\int_0^1\frac{d\Gamma}{d\hat{s}dz}}(z,\hat s,1/R)\,dz -
\ds{\int_{-1}^0\frac{d\Gamma}{d\hat{s}dz}}(z,\hat s,1/R)\,dz}
{\ds{\int_0^1\frac{d\Gamma}{d\hat{s}dz}}(z,\hat s,1/R)\,dz +
\ds{\int_{-1}^0\frac{d\Gamma}{d\hat{s}dz}}(z,\hat s,1/R)\,dz}~. \eea
The dependence of forward-backward asymmetry on $1/R$ for the decay under consideration in both lepton channels is depicted in Figure 3.
\begin{figure}[h!]
\centering
\begin{tabular}{ccc}
\epsfig{file=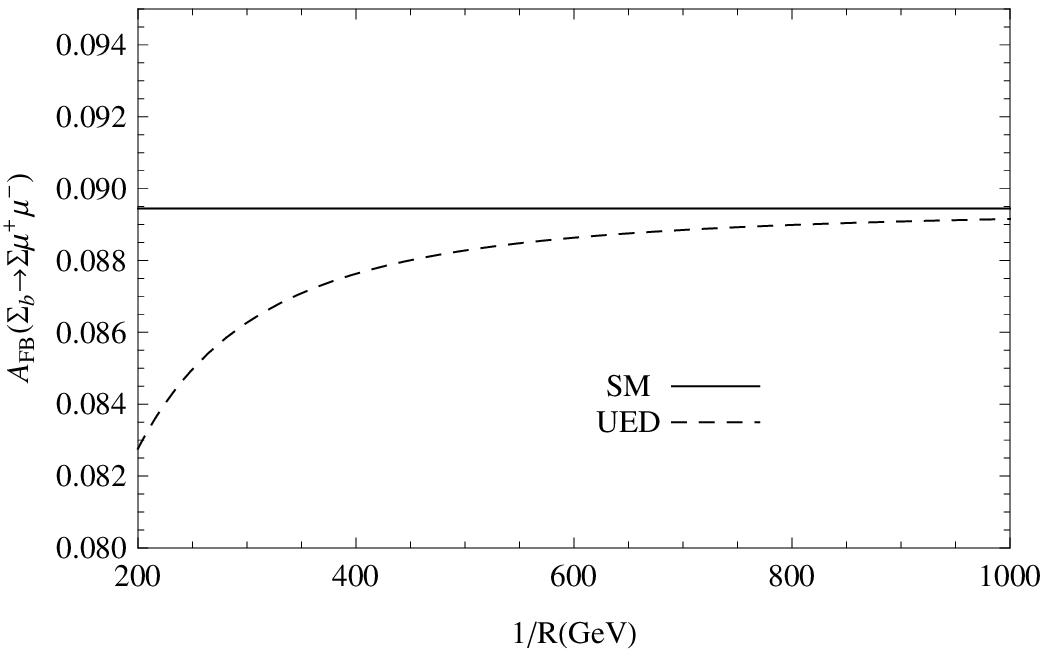,width=0.45\linewidth,clip=} &
\epsfig{file=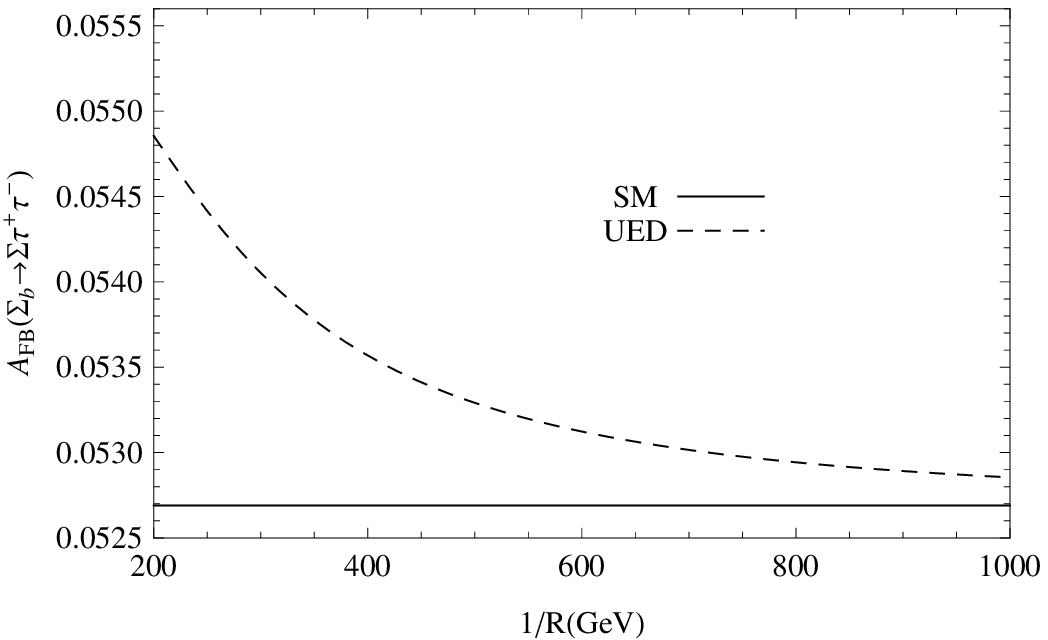,width=0.45\linewidth,clip=}
\end{tabular}
\caption{The $1/R$ dependence of forward-backward asymmetry for $\Sigma_{b}\rightarrow \Sigma \ell^{+}\ell^{-}$ at $\hat s=0.5$.}
\end{figure}
With a glance in this figure, we read
\begin{itemize}
 \item there are also considerable discrepancies between two models predictions in both lepton channels at small values of $1/R$.
 \item  As far as the $\mu$ channel is concerned, the values obtained in UED at lower values of compactification scale are small compared to the SM predictions. In $\tau$ channel,
we have inverse situation.
\end{itemize}

\subsection{ $\Sigma$ Baryon Polarizations}
In this part we deal with the $\Sigma$ baryon polarizations.
 To define these polarizations,
we write the $\Sigma$ baryon spin four--vector in terms of a unit vector
$\vec{\xi}$ along the $\Sigma$ baryon spin in its rest frame (for more details see \cite{baryon-pol1,ozpineci,baryon-pol2}), i.e.,  
\bea
\label{e14} 
s_\mu = \ga \frac{\vec{p}_\Sigma \cdot \vec{\xi}}{m_\Sigma},
\vec{\xi} + \frac{\vec{p}_\Sigma (\vec{p}_\Sigma \cdot
\vec{\xi})}{E_\Sigma+m_\Sigma} \dr ~,
\eea   
and select the following unit vectors along the longitudinal, transversal and normal
components:
\bea
\label{e15}
\vec{e}_L = \frac{\vec{p}_\Sigma}{\vel \vec{p}_\Sigma \ver}~, ~~~
\vec{e}_T = \frac{\vec{p}_\ell\times \vec{p}_\Sigma}
{\vel \vec{p}_\ell\times \vec{p}_\Sigma \ver}~,~~~
\vec{e}_N = \vec{e}_T \times \vec{e}_L~,
\eea
 where  $\vec{p}_\ell$ and $\vec{p}_\Sigma$ are the three
momenta of $\ell$ lepton and $\Sigma$ baryon, in the center of mass frame of the 
$\ell^+ \ell^-$.
The $1/R$-dependent differential decay rate of the $\Sigma_b \rar \Sigma \ell^+ \ell^-$ transition
for any spin direction $\vec{\xi}$ along the $\Sigma$ baryon
can be written as
\bea
\label{e16}
\frac{d\Gamma(\vec{\xi})}{d\hat s} (\hat s,1/R)= \frac{1}{2}
\ga \frac{d\Gamma}{d\hat s}(\hat s,1/R)\dr
\Bigg[ 1 + \Bigg( P_L(\hat s,1/R) \vec{e}_L + P_N(\hat s,1/R)
\vec{e}_N + P_T (\hat s,1/R)\vec{e}_T \Bigg) \cdot
\vec{\xi} \Bigg]~,\nnb\\
\eea
where, the  $  \frac{d\Gamma}{d\hat s}(\hat s,1/R) $ in right hand side is the differential decay rate   corresponds to the unpolarized case defined at Eq.(\ref{decayrate}). The    
$P_L$, $P_N$ and $P_T$ in the above equation stand for the longitudinal, normal and 
transversal polarizations of the $\Sigma$ baryon, respectively. 
They are defined as:
\bea
\label{e18}
P_i(q^2) = \frac{\ds{\frac{d \Gamma}{d\hat s}
                   (\vec{\xi}=\vec{e}_i) -
                   \frac{d \Gamma}{d\hat s}
                   (\vec{\xi}=-\vec{e}_i)}}
              {\ds{\frac{d \Gamma}{d\hat s}
                   (\vec{\xi}=\vec{e}_i) +
                  \frac{d \Gamma}{d\hat s}
                  (\vec{\xi}=-\vec{e}_i)}}~,
\eea
where $i = L,N$ or $T$. These definitions lead to the following explicit expressions of the $\Sigma$ baryon polarizations:
\bea \label{PL} P_L (\hat s, 1/R)\es
\frac{16 m_{\Sigma_b}^2 \sqrt{\lambda}}{\Delta(\hat s, 1/R)} \Bigg\{ 8 m_\ell^2
m_{\Sigma_b}\, \Big( \mbox{\rm Re}[D_1^\ast E_3 - D_3^\ast E_1] +
\sqrt{r} \mbox{\rm Re}[D_1^\ast D_3 - E_1^\ast E_3)] \Big) \nnb \\
\ar 2 m_\ell m_{\Sigma_b}\,  (1+\sqrt{r}) \mbox{\rm
Re}[(D_1-E_1)^\ast F_2]
 \nnb \\
\ek 2 m_\ell m_{\Sigma_b}^2 \hat s \, \Big\{ \mbox{\rm
Re}[(D_3-E_3)^\ast F_2 ] +
2 m_\ell ( \vel D_3 \ver^2 - \vel E_3 \ver^2 ) \Big\} \nnb \\
\ek 4 m_{\Sigma_b} (2 m_\ell^2 + m_{\Sigma_b}^2 \hat s) \,
\mbox{\rm Re}[A_1^\ast B_2 - A_2^\ast B_1] \nnb \\
\ek \frac{4}{3} m_{\Sigma_b}^3 \hat s v^2 \, \Big( 3 \mbox{\rm
Re}[D_1^\ast E_2 - D_2^\ast E_1] +
\sqrt{r} \mbox{\rm Re}[D_1^\ast D_2 - E_1^\ast E_2] \Big) \nnb \\
\ek \frac{4}{3} m_{\Sigma_b} \sqrt{r} (6 m_\ell^2 +
m_{\Sigma_b}^2 \hat s v^2) \, \mbox{\rm Re}[A_1^\ast A_2 - B_1^\ast B_2] \nnb \\
\ar \frac{1}{3} \Big\{ 3 [4 m_\ell^2 + m_{\Sigma_b}^2 (1-r+\hat s)]
(\vel A_1 \ver^2 -
\vel B_1 \ver^2 ) - 3 [4 m_\ell^2 -  m_{\Sigma_b}^2 (1-r+\hat s)] \nnb \\
\cp (\vel D_1 \ver^2 - \vel E_1 \ver^2 ) -  m_{\Sigma_b}^2
(1-r-\hat s) v^2 (\vel A_1 \ver^2 - \vel B_1 \ver^2 + \vel D_1 \ver^2 -
\vel E_1 \ver^2 )
\Big\} \nnb \\
\ek \frac{1}{3} m_{\Sigma_b}^2 \{ 12 m_\ell^2 (1-r) +
m_{\Sigma_b}^2 \hat s [3 (1-r+\hat s) + v^2 (1-r-\hat s)] \}
(\vel A_2 \ver^2 - \vel B_2 \ver^2) \nnb \\
\ek \frac{2}{3} m_{\Sigma_b}^4 \hat s (2 - 2 r + \hat s) v^2 \,
(\vel D_2 \ver^2 - \vel E_2 \ver^2)
\Bigg\}~,  \eea

\bea \label{PN} P_N (\hat s, 1/R)\es \frac{8 \pi m_{\Sigma_b}^3 v
\sqrt{\hat s}}{\Delta (\hat s, 1/R)}\Bigg\{
- 2 m_{\Sigma_b} (1-r+\hat s) \sqrt{r} \,
\mbox{\rm Re}[A_1^\ast D_1 + B_1^\ast E_1] \nnb \\
\ar m_{\Sigma_b} (1-\sqrt{r}) [(1+\sqrt{r})^2 -\hat s] \, \Big(
 m_\ell \mbox{\rm Re}[(A_2-B_2)^\ast F_1] \Big) \nnb \\
\ar m_\ell [(1+\sqrt{r})^2 -\hat s] \,
\mbox{\rm Re}[A_1^\ast F_1] \nnb \\
\ar 4 m_{\Sigma_b}^2 \hat s \sqrt{r} \, \mbox{\rm Re}[A_1^\ast E_2 +
A_2^\ast E_1 +B_1^\ast D_2 +
B_2^\ast D_1] \nnb \\
\ek 2 m_{\Sigma_b}^3 \hat s \sqrt{r} (1-r+\hat s) \,
\mbox{\rm Re}[A_2^\ast D_2 + B_2^\ast E_2^\ast] \nnb \\
\ar 2 m_{\Sigma_b} (1-r-\hat s) \, \Big( \mbox{\rm Re}[A_1^\ast E_1 +
B_1^\ast D_1] +
m_{\Sigma_b}^2 \hat s \mbox{\rm Re}[A_2^\ast E_2 + B_2^\ast D_2] \Big) \nnb \\
\ek m_{\Sigma_b}^2 [(1-r)^2-\hat s^2] \, \mbox{\rm Re}[A_1^\ast D_2 +
A_2^\ast D_1 + B_1^\ast E_2 +
B_2^\ast E_1] \nnb \\
\ek m_\ell [(1+\sqrt{r})^2 -\hat s] \,
\mbox{\rm Re}[B_1^\ast F_1]
\Bigg\}~ , \eea
 \bea
\label{e12}
\lefteqn{
P_T(\hat s, 1/R) = - \frac{8 \pi m_{\Sigma_b}^3 v \sqrt{\hat s\lambda}}
{\Delta(\hat s, 1/R)}
\Bigg\{
m_\ell \Big(
\mbox{\rm Im}[(A_1+B_1)^\ast F_1]
\Big)} \nnb \\
\ek m_\ell m_{\Sigma_b} \Big[
(1+\sqrt{r}) \, \mbox{\rm Im}[(A_2+B_2)^\ast F_1] \Big] \nnb \\
\ar m_{\Sigma_b}^2 (1-r+\hat s) \Big(
\mbox{\rm Im}[A_2^\ast D_1 - A_1^\ast D_2] -
\mbox{\rm Im}[B_2^\ast E_1 - B_1^\ast E_2] \Big)\nnb \\
\ar 2 m_{\Sigma_b} \Big( \mbox{\rm Im}[A_1^\ast E_1-B_1^\ast D_1]
- m_{\Sigma_b}^2 \hat s \, \mbox{\rm Im}[A_2^\ast E_2 - B_2^\ast
D_2] \Big)\Bigg\}~, \eea
where \bea \Delta(\hat s, 1/R)\es{{\cal
T}_0(\hat s, 1/R) +\frac{1}{3} {\cal T}_2(\hat s, 1/R)},
 \eea
and $F_1=g_1-\sqrt r m_{\Sigma_b} g_2$, $ F_2= m_{\Sigma_b} g_2$.

The dependence of different $\Sigma$ baryon polarizations on compactification scale at  $\mu$ and $\tau$ channels in the SM and UED are presented in Figures 4-6.  
\begin{figure}[h!]
\centering
\begin{tabular}{ccc}
\epsfig{file=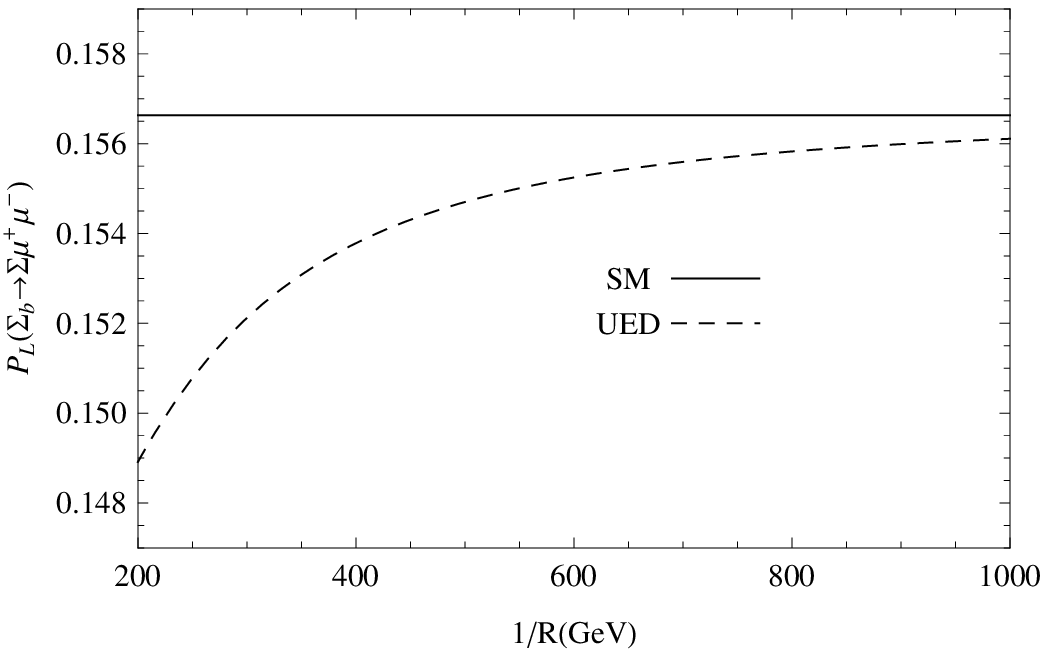,width=0.45\linewidth,clip=} &
\epsfig{file=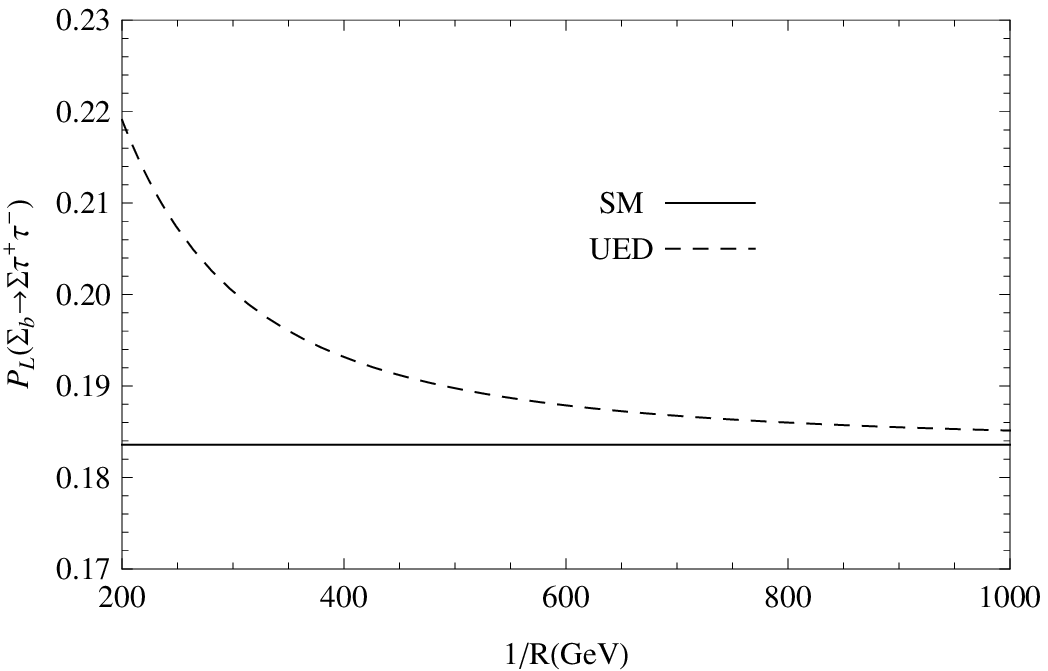,width=0.45\linewidth,clip=}
\end{tabular}
\caption{The $1/R$ dependence of the longitudinal polarization, $P_L(\hat s,1/R)$ at $\hat s=0.5$.}
\end{figure}
\begin{figure}[h!]
\centering
\begin{tabular}{ccc}
\epsfig{file=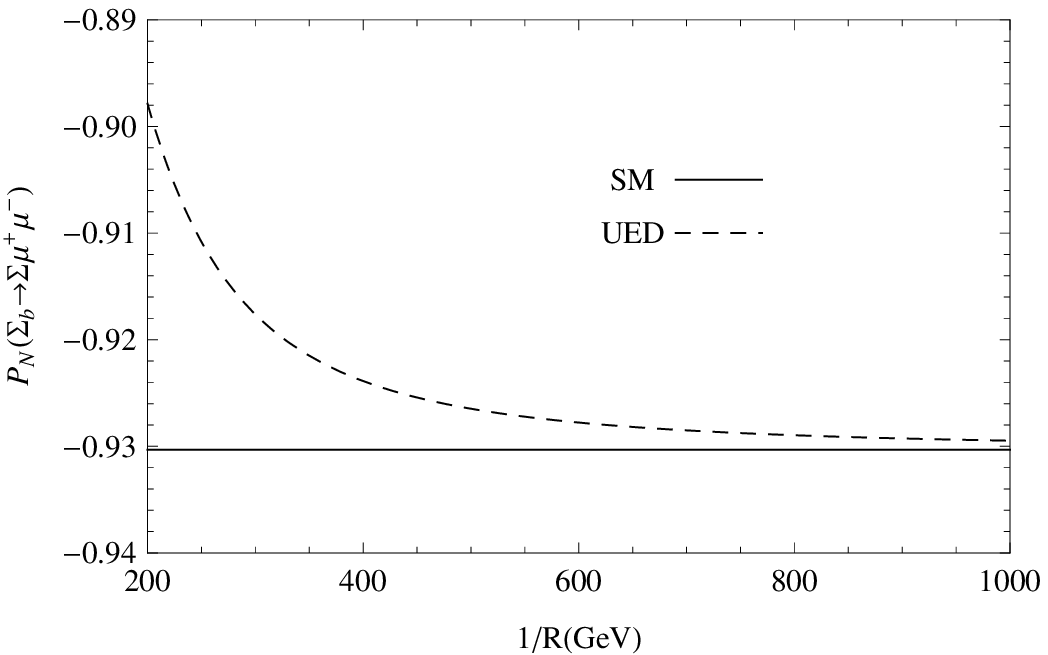,width=0.45\linewidth,clip=} &
\epsfig{file=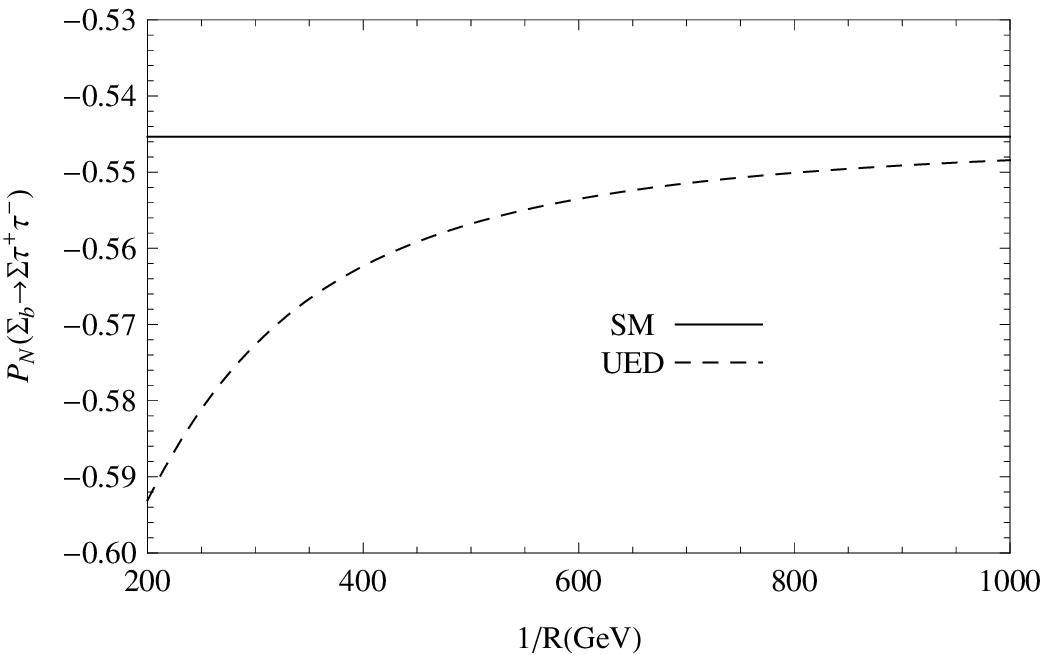,width=0.45\linewidth,clip=}
\end{tabular}
\caption{The $1/R$ dependence of the normal polarization, $P_N(\hat s,1/R)$ at $\hat s=0.5$.}
\end{figure}
\begin{figure}[h!]
\centering
\begin{tabular}{ccc}
\epsfig{file=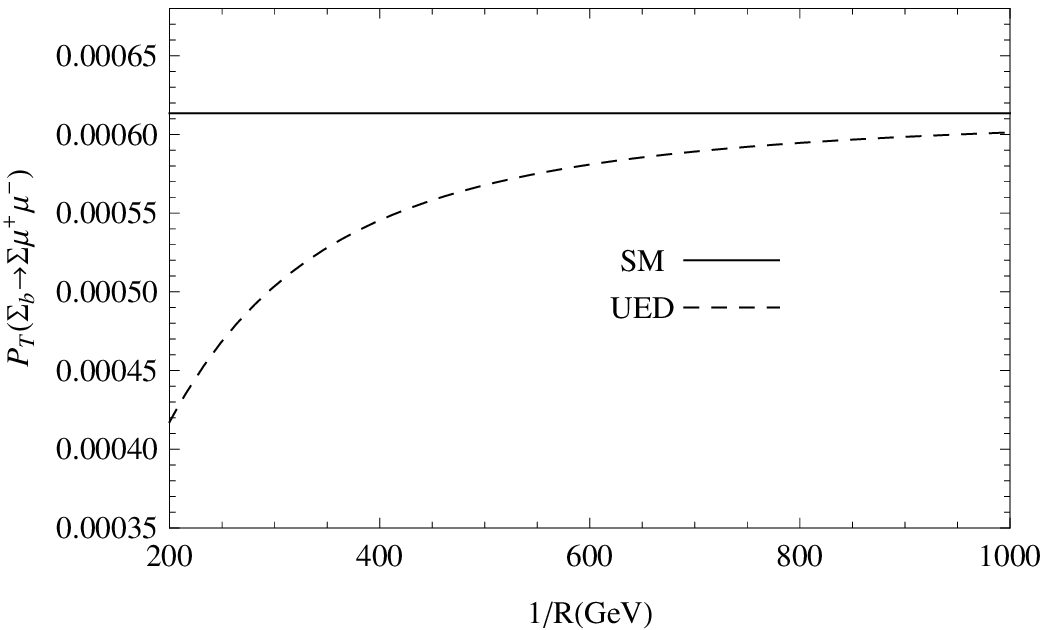,width=0.45\linewidth,clip=} &
\epsfig{file=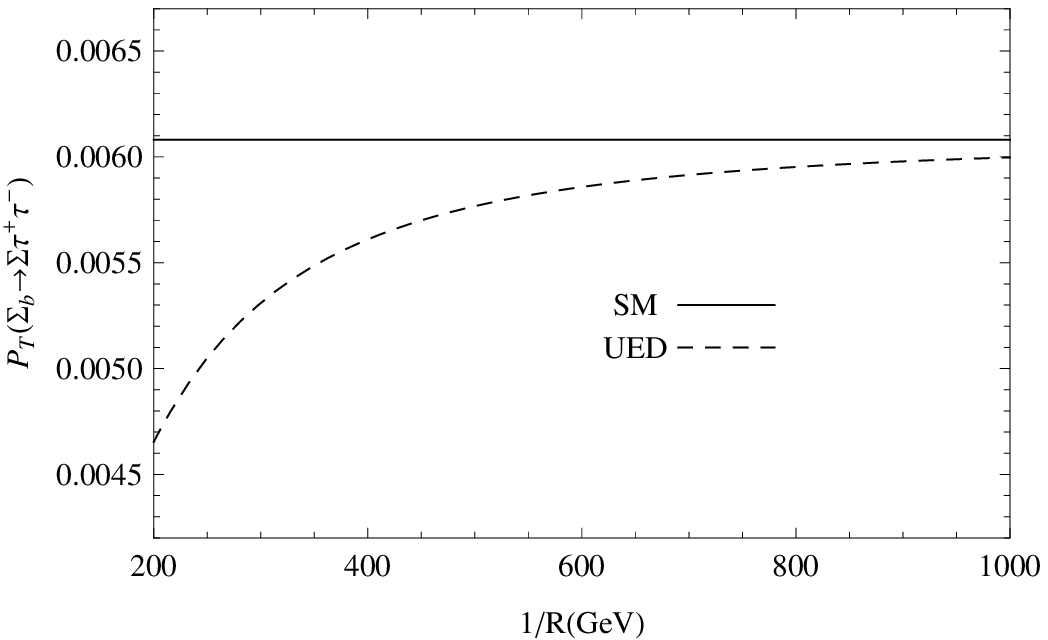,width=0.45\linewidth,clip=}
\end{tabular}
\caption{The $1/R$ dependence of the transversal polarization, $P_T(\hat s,1/R)$ at $\hat s=0.5$.}
\end{figure}
From these figures, we conclude that
 \begin{itemize}
\item the UED predictions deviate considerably from those of the SM for all polarizations and both lepton channels at small values of compactification scale.
\item The numerical values show that the $P_L$ and $P_N$ have measurable sizes for both leptons but $P_T$ is very small.
 \item In the case of $P_L$ and $|P_N|$, the UED predictions at lower values of  $1/R$ are smaller than those of the SM at $\mu$ channel. However, for  $\tau$ we have inverse situation. In the case
of $P_T$, two lepton channels represent similar behavior.
 \end{itemize}

\subsection{Double Lepton Polarization Asymmetries}
The present subsection encompasses our analysis on   the double--lepton polarization asymmetries. In the case of  both leptons polarizations, 
we define the following orthogonal unit vectors
$s_i^{\pm\mu}$ with again  $i=L,T$ or $N$ in the rest frame of double leptons (For details see for instance \cite{Aliev-Sirvanli,aliev-bashiry,Bensalem}):
\bea
\label{e6616}   
s^{-\mu}_L \es \ga 0,\vec{e}_L^{\,-}\dr =
\ga 0,\frac{\vec{p}_-}{\vel\vec{p}_- \ver}\dr~, \nnb \\
s^{-\mu}_N \es \ga 0,\vec{e}_N^{\,-}\dr = \ga 0,\frac{\vec{p}_\Sigma\times
\vec{p}_-}{\vel \vec{p}_\Sigma\times \vec{p}_- \ver}\dr~, \nnb \\
s^{-\mu}_T \es \ga 0,\vec{e}_T^{\,-}\dr = \ga 0,\vec{e}_N^{\,-}
\times \vec{e}_L^{\,-} \dr~, \nnb \\
s^{+\mu}_L \es \ga 0,\vec{e}_L^{\,+}\dr =
\ga 0,\frac{\vec{p}_+}{\vel\vec{p}_+ \ver}\dr~, \nnb \\
s^{+\mu}_N \es \ga 0,\vec{e}_N^{\,+}\dr = \ga 0,\frac{\vec{p}_\Sigma\times
\vec{p}_+}{\vel \vec{p}_\Sigma\times \vec{p}_+ \ver}\dr~, \nnb \\
s^{+\mu}_T \es \ga 0,\vec{e}_T^{\,+}\dr = \ga 0,\vec{e}_N^{\,+}
\times \vec{e}_L^{\,+}\dr~,
\eea
where $\vec{p}_\pm$ and $\vec{p}_\Sigma$ are the three--momenta of the
leptons $\ell^\pm$ and $\Sigma$ baryon. Now, by the help of the Lorentz boost, we transform these unit vectors from the rest frame of the leptons to center of mass (CM)
frame of them along the longitudinal direction. As a result for the unit vectors $s_L^{\pm\mu}$ we get
\bea
\label{e6617}
\ga s^{\mp\mu}_L \dr_{CM} \es \ga \frac{\vel\vec{p}_\mp \ver}{m_\ell}~,
\frac{E_\ell \vec{p}_\mp}{m_\ell \vel\vec{p}_\mp \ver}\dr~,
\eea
where, $\vec{p}_+ = - \vec{p}_-$; and  $E_\ell$ and $m_\ell$ are the energy and mass
of leptons in the CM frame, respectively. The remaining two unit vectors, $s_N^{\pm\mu}$, $s_T^{\pm\mu}$ do not change
under the considered transformation. We now define the double--polarization 
asymmetries as:

\bea                                                                  
\label{e6618}
P_{ij}(\hat s,1/R) \es
\frac{ 
\Big( \ds \frac{d\Gamma(\vec{s}^{\,-}_i,\vec{s}^{\,+}_j)}{d\hat s} -
      \ds \frac{d\Gamma(-\vec{s}^{\,-}_i,\vec{s}^{\,+}_j)}{d\hat s} \Big) -
\Big( \ds \frac{d\Gamma(\vec{s}^{\,-}_i,-\vec{s}^{\,+}_j)}{d\hat s} -      
      \ds \frac{d\Gamma(-\vec{s}^{\,-}_i,-\vec{s}^{\,+}_j)}{d\hat s} \Big) 
     }
     {    
\Big( \ds \frac{d\Gamma(\vec{s}^{\,-}_i,\vec{s}^{\,+}_j)}{d\hat s} +      
      \ds \frac{d\Gamma(-\vec{s}^{\,-}_i,\vec{s}^{\,+}_j)}{d\hat s} \Big) +
\Big( \ds \frac{d\Gamma(\vec{s}^{\,-}_i,-\vec{s}^{\,+}_j)}{d\hat s} +      
      \ds \frac{d\Gamma(-\vec{s}^{\,-}_i,-\vec{s}^{\,+}_j)}{d\hat s} \Big)
     }~.
\eea
Using this definition, we obtain the following $1/R$-dependent expressions for the double lepton polarization asymmetries :
\bea \label{PLN} P_{LN}(\hat
s, 1/R) \es \frac{16 \pi m_{\Sigma_b}^4 \hat{m}_\ell
\sqrt{\lambda}}{\Delta(\hat s, 1/R) \sqrt{\hat{s}}} \mbox{\rm Im}
\Bigg\{
(1-r) (A_1^\ast D_1 + B_1^\ast E_1)
+ m_{\Sigma_b}
 \hat{s} (A_1^\ast E_3 - A_2^\ast E_1 + B_1^\ast D_3
-B_2^\ast D_1) \nnb \\
\ar m_{\Sigma_b}
 \sqrt{r} \hat{s}
(A_1^\ast D_3 + A_2^\ast D_1 +B_1^\ast E_3 + B_2^\ast E_1)
- m_{\Sigma_b}^2 \hat{s}^2 \Big( B_2^\ast E_3 + A_2^\ast D_3
\Big) \Bigg\},\eea

\begin{figure}[h!]
\centering
\begin{tabular}{ccc}
\epsfig{file=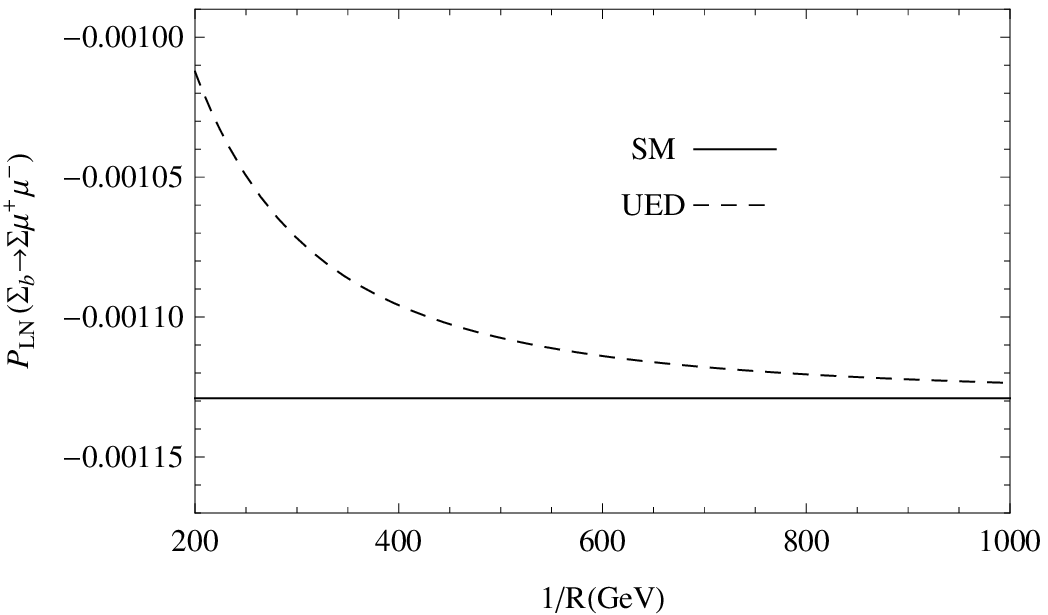,width=0.45\linewidth,clip=} &
\epsfig{file=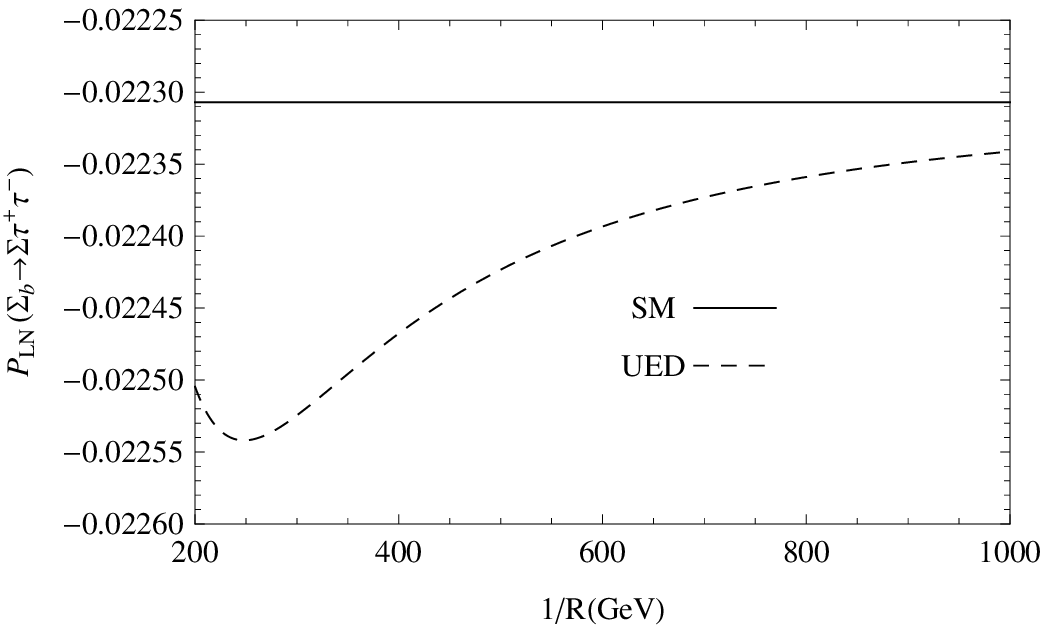,width=0.45\linewidth,clip=}
\end{tabular}
\caption{The $1/R$ dependence of the $P_{LN}(\hat s,1/R)$ at $\hat s=0.5$.}
\end{figure}

\bea \label{PLT} P_{LT}(\hat s, 1/R) \es \frac{16 \pi m_{\Sigma_b}^4
\hat{m}_\ell \sqrt{\lambda} v}{\Delta(\hat s, 1/R) \sqrt{\hat{s}}} \mbox{\rm
Re} \Bigg\{
(1-r) \Big( \vel D_1 \ver^2 + \vel E_1 \ver^2 \Big)
- \hat{s} \Big(A_1 D_1^\ast - B_1 E_1^\ast \Big) \nnb \\
\ek m_{\Sigma_b} \hat{s} \Big[ B_1 D_2^\ast + (A_2 + D_2 -D_3)
E_1^\ast -  A_1 E_2^\ast
-(B_2-E_2+E_3) D_1^\ast \Big] \nnb \\
\ar m_{\Sigma_b}
 \sqrt{r} \hat{s}
\Big[ A_1 D_2^\ast + (A_2 + D_2 +D_3) D_1^\ast - B_1 E_2^\ast -
(B_2 - E_2 - E_3) E_1^\ast \Big] \nnb \\
\ar m_{\Sigma_b}^2 \hat{s} (1-r) (A_2 D_2^\ast -
B_2 E_2^\ast)
- m_{\Sigma_b}^2 \hat{s}^2 (D_2 D_3^\ast + E_2 E_3^\ast )\Bigg\}, \eea

\bea \label{e7716} P_{LL}(\hat s, 1/R) \es \frac{16 m_{\Sigma_b}^4}{3\Delta(\hat s, 1/R)}
\mbox{\rm Re} \Bigg\{\nnb \\
\ek 6 m_{\Sigma_b} \sqrt{r}
(1-r+\hat{s}) \Big[ \hat{s} (1+v^2) (A_1 A_2^\ast +
B_1 B_2^\ast)  -
4 \hat{m}_\ell^2 (D_1 D_3^\ast + E_1 E_3^\ast) \Big] \nnb \\
\ar 6 m_{\Sigma_b} (1-r-\hat{s}) \Big[ \hat{s}
(1+v^2) (A_1 B_2^\ast + A_2 B_1^\ast) +
4 \hat{m}_\ell^2 (D_1 E_3^\ast + D_3 E_1^\ast) \Big] \nnb \\
\ar 12 \sqrt{r} \hat{s} (1+v^2) \Big( A_1 B_1^\ast +
D_1 E_1^\ast +
m_{\Sigma_b}^2 \hat{s} A_2 B_2^\ast \Big) \nnb \\
\ar 12 m_{\Sigma_b}^2 \hat{m}_\ell^2 \hat{s}
(1+r-\hat{s})
\ga \vel D_3 \ver^2 + \vel E_3^\ast \ver^2 \dr \nnb \\
\ek (1+v^2) \Big[ 1+r^2 - r
(2-\hat{s}) +\hat{s} (1-2 \hat{s}) \Big]
\Big(\vel A_1 \ver^2 + \vel B_1 \ver^2 \Big) \nnb \\
\ek \Big[ (5 v^2 - 3) (1-r)^2 + 4 \hat{m}_\ell^2
(1+r) + 2 \hat{s} (1+8 \hat{m}_\ell^2 +
r)
- 4 \hat{s}^2 \Big] \Big( \vel D_1 \ver^2 + \vel E_1 \ver^2 \Big) \nnb \\
\ek m_{\Sigma_b}^2 (1+v^2) \hat{s} \Big[2 + 2 r^2
-\hat{s}(1 +\hat{s}) - r (4 + \hat{s})\Big] \big(
\vel A_2 \ver^2 + \vel B_2 \ver^2 \Big) \nnb \\
\ek 2 m_{\Sigma_b}^2 \hat{s} v^2 \Big[ 2 (1 + r^2)
- \hat{s} (1+\hat{s}) - r (4+\hat{s})\Big] \Big(
\vel D_2 \ver^2 + \vel E_2 \ver^2 \Big) \nnb \\
\ar 12 m_{\Sigma_b} \hat{s} (1-r-\hat{s}) v^2
\Big( D_1 E_2^\ast + D_2 E_1^\ast \Big) \nnb \\
\ek 12 m_{\Sigma_b} \sqrt{r} \hat{s}
(1-r+\hat{s}) v^2
\Big( D_1 D_2^\ast + E_1 E_2^\ast \Big) \nnb \\
\ar 24 m_{\Sigma_b}^2 \sqrt{r} \hat{s} \Big(
\hat{s} v^2 D_2 E_2^\ast + 2 \hat{m}_\ell^2 D_3 E_3^\ast \Big)\Bigg\},
\eea

\bea \label{PNL} P_{NL} (\hat s, 1/R)\es - \frac{16 \pi m_{\Sigma_b}^4
\hat{m}_\ell \sqrt{\lambda}}{\Delta(\hat s, 1/R) \sqrt{\hat{s}}} \mbox{\rm Im}
\Bigg\{
(1-\hat{r}_\Sigma) (A_1^\ast D_1 + B_1^\ast E_1)
+ m_{\Sigma_b}
 \hat{s} (A_1^\ast E_3 - A_2^\ast E_1 + B_1^\ast D_3
-B_2^\ast D_1) \nnb \\
\ek m_{\Sigma_b}
 \sqrt{\hat{r}_\Sigma} \hat{s}
(A_1^\ast D_3 + A_2^\ast D_1 +B_1^\ast E_3 + B_2^\ast E_1)
- m_{\Sigma_b}^2 \hat{s}^2 \Big( B_2^\ast E_3 + A_2^\ast D_3
\Big) \Bigg\},\eea
\begin{figure}[h!]
\centering
\begin{tabular}{ccc}
\epsfig{file=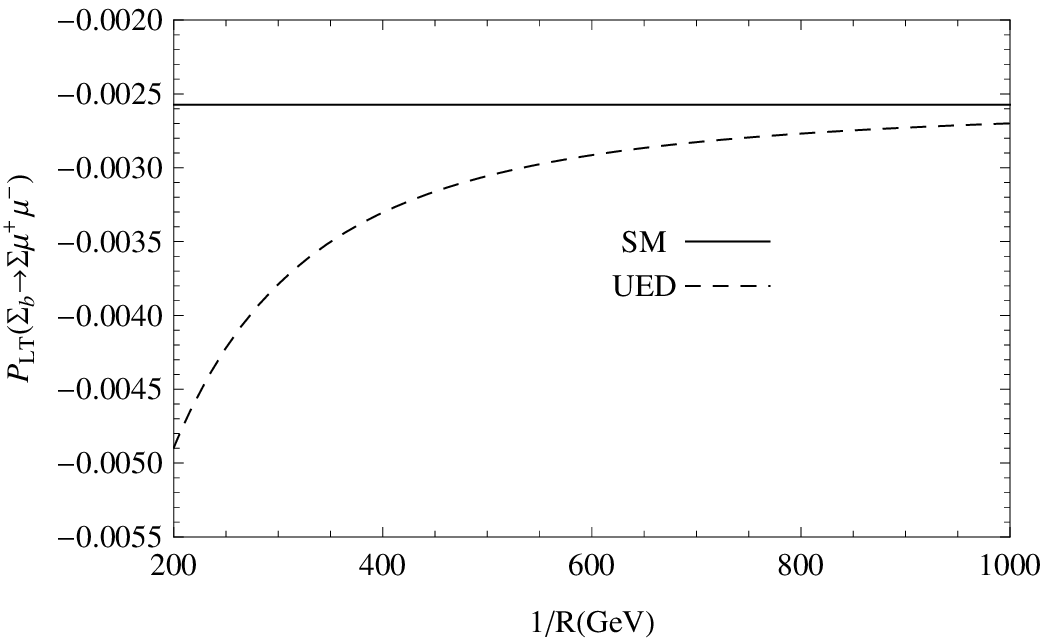,width=0.45\linewidth,clip=} &
\epsfig{file=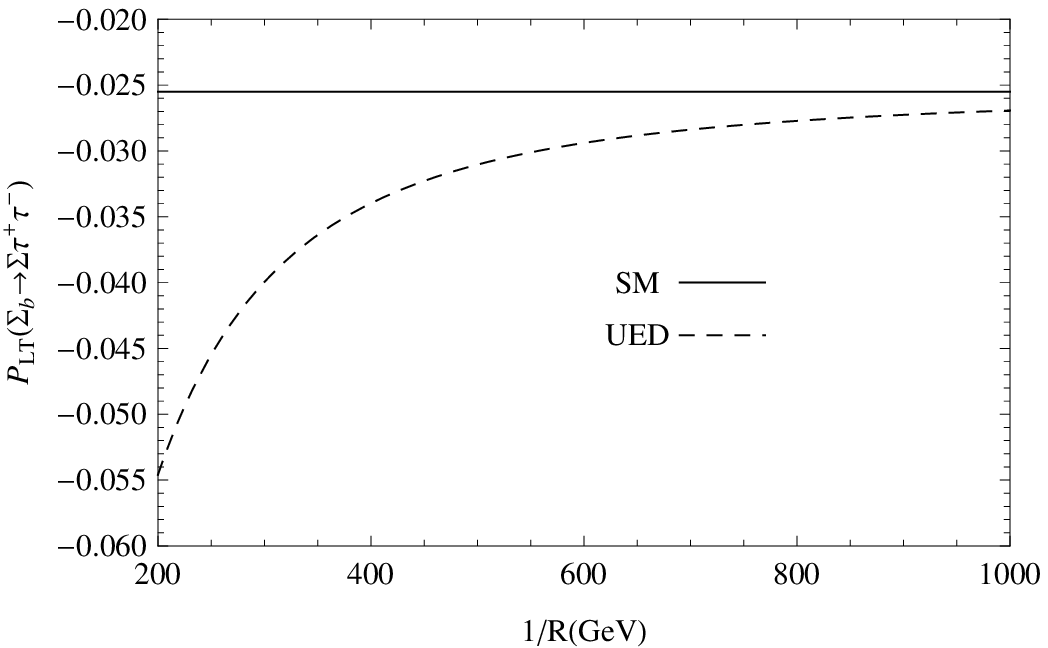,width=0.45\linewidth,clip=}
\end{tabular}
\caption{The $1/R$ dependence of the $P_{LT}(\hat s,1/R)$ at $\hat s=0.5$.}
\end{figure}
\begin{figure}[h!]
\centering
\begin{tabular}{ccc}
\epsfig{file=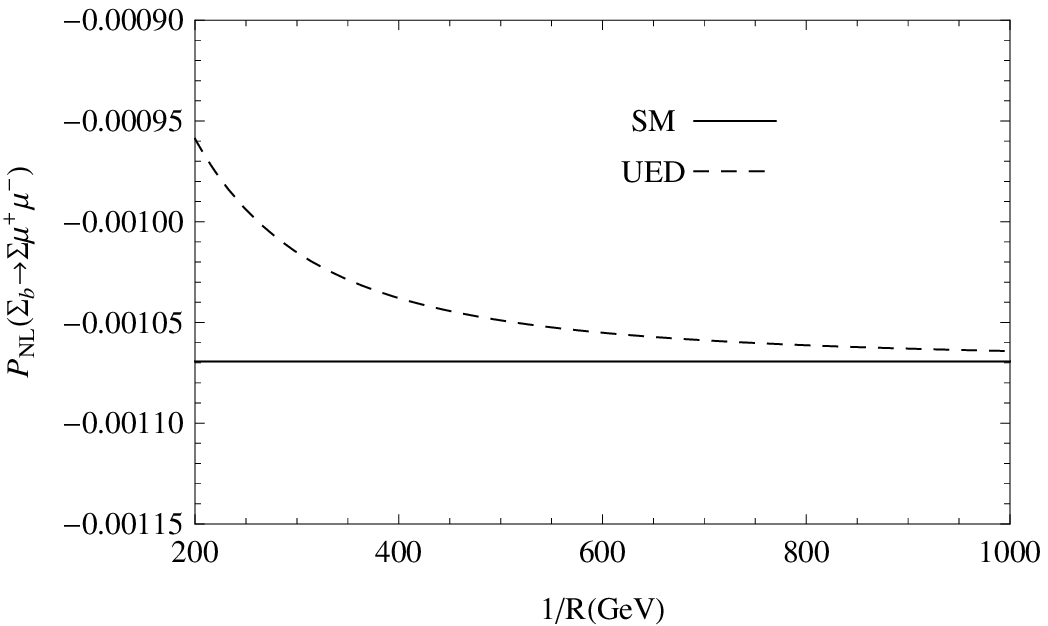,width=0.45\linewidth,clip=} &
\epsfig{file=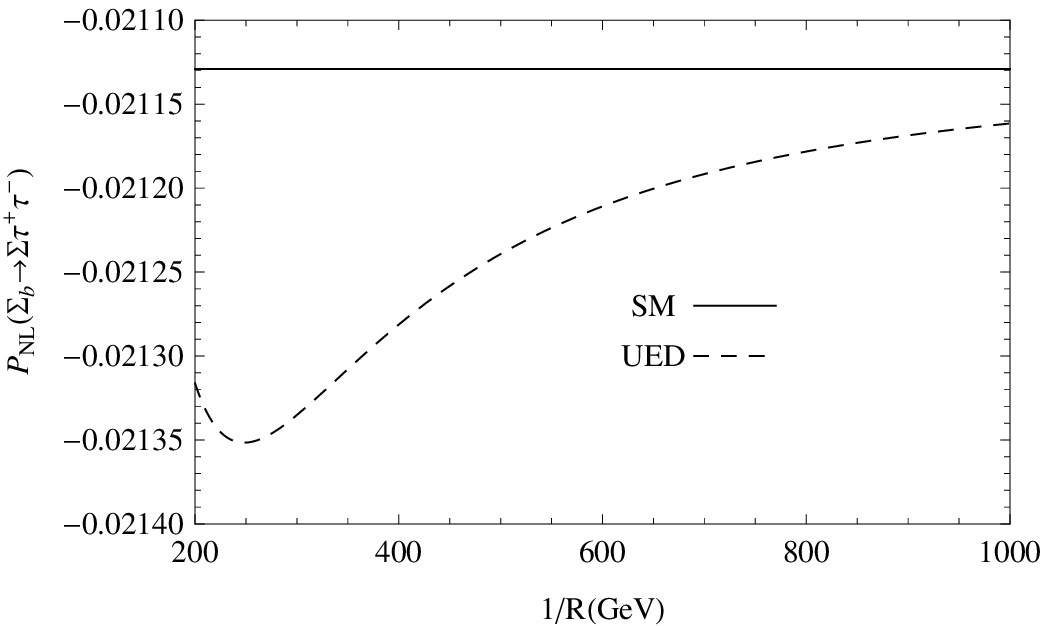,width=0.45\linewidth,clip=}
\end{tabular}
\caption{The $1/R$ dependence of the $P_{NL}(\hat s,1/R)$ at $\hat s=0.5$.}
\end{figure}

\bea \label{PNN} P_{NN} (\hat s, 1/R)\es \frac{32 m_{\Sigma_b}^4}{3 \hat{s}
\Delta(\hat s, 1/R)} \mbox{\rm Re} \Bigg\{
24 \hat{m}_\ell^2 \sqrt{r} \hat{s}
( A_1 B_1^\ast + D_1 E_1^\ast ) \nnb \\
\ek 12 m_{\Sigma_b} \hat{m}_\ell^2 \sqrt{r} \hat{s}
(1-r +\hat{s}) (A_1 A_2^\ast + B_1 B_2^\ast) \nnb \\
\ar 6 m_{\Sigma_b} \hat{m}_\ell^2 \hat{s} \Big[ m_{\Sigma_b}
\hat{s} (1+r-\hat{s}) \Big(\vel D_3 \ver^2 + \vel
E_3 \ver^2 \Big) + 2 \sqrt{r}
(1-r+\hat{s})
(D_1 D_3^\ast + E_1 E_3^\ast)\Big] \nnb \\
\ar 12 m_{\Sigma_b} \hat{m}_\ell^2 \hat{s}
(1-r-\hat{s})
(A_1 B_2^\ast + A_2 B_1 ^\ast + D_1 E_3^\ast + D_3 E_1^\ast) \nnb \\
\ek [ \lambda \hat{s} + 2 \hat{m}_\ell^2 (1 + r^2 -
2 r + r \hat{s} + \hat{s} - 2
\hat{s}^2) ] \Big( \vel A_1 \ver^2 + \vel B_1 \ver^2 - \vel D_1
\ver^2 -
\vel E_1 \ver^2 \Big) \nnb \\
\ar 24 m_{\Sigma_b}^2 \hat{m}_\ell^2 \sqrt{r}
\hat{s}^2 (A_2 B_2^\ast + D_3 E_3^\ast)
- m_{\Sigma_b}^2 \lambda \hat{s}^2 v^2
\Big( \vel D_2 \ver^2 + \vel E_2 \ver^2 \Big) \nnb \\
\ar m_{\Sigma_b}^2 \hat{s} \{ \lambda \hat{s} - 2 \hat{m}_\ell^2
[2 (1+ r^2) - \hat{s} (1+\hat{s}) - r
(4+\hat{s})]\} \Big( \vel A_2 \ver^2 + \vel B_2 \ver^2 \Big)\Bigg\},
\eea
\begin{figure}[h!]
\centering
\begin{tabular}{ccc}
\epsfig{file=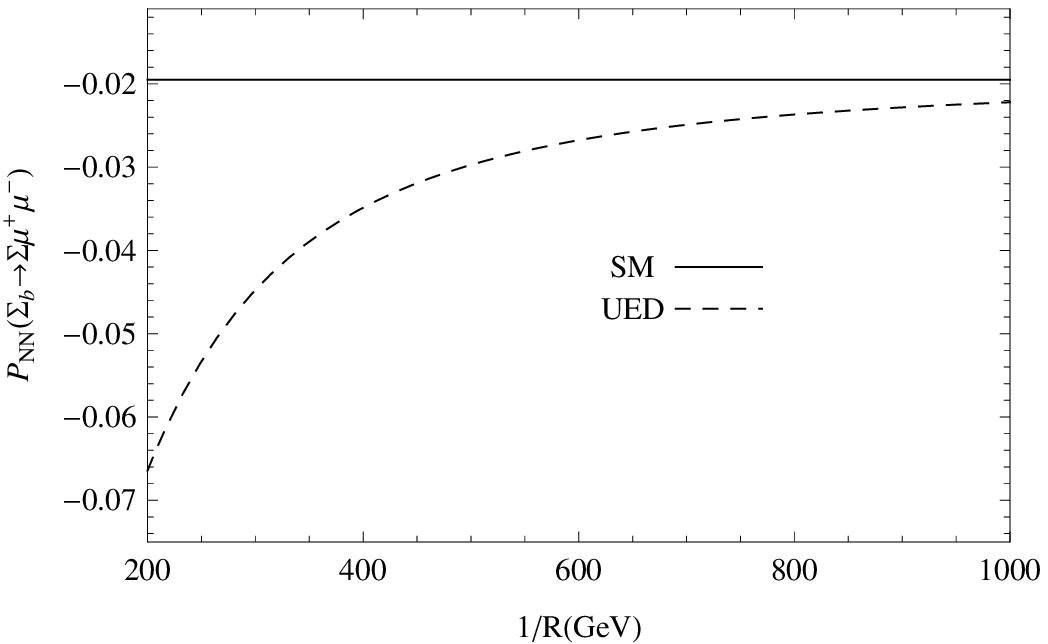,width=0.45\linewidth,clip=} &
\epsfig{file=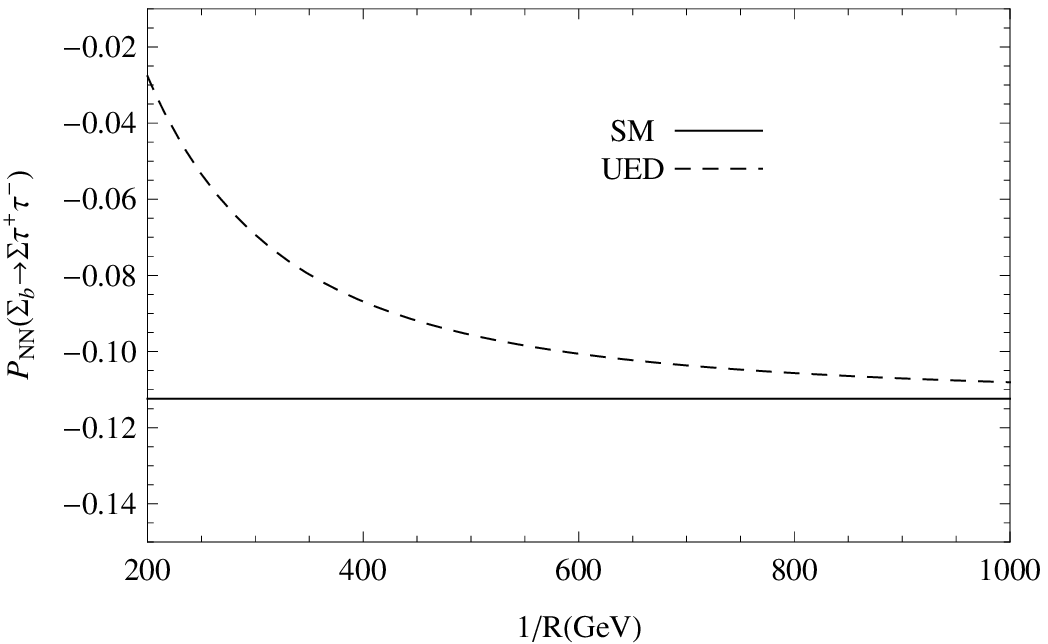,width=0.45\linewidth,clip=}
\end{tabular}
\caption{The $1/R$ dependence of the $P_{NN}(\hat s,1/R)$ at $\hat s=0.5$.}
\end{figure}
\bea \label{PNT} P_{NT} (\hat s, 1/R)\es \frac{64 m_{\Sigma_b}^4 \lambda
v}{3 \Delta(\hat s, 1/R)} \mbox{\rm Im} \Bigg\{
(A_1 D_1^\ast +B_1 E_1^\ast)
+ m_{\Sigma_b}^2 \hat{s} (A_2^\ast D_2 + B_2^\ast E_2)\Bigg\},
\eea
\begin{figure}[h!]
\centering
\begin{tabular}{ccc}
\epsfig{file=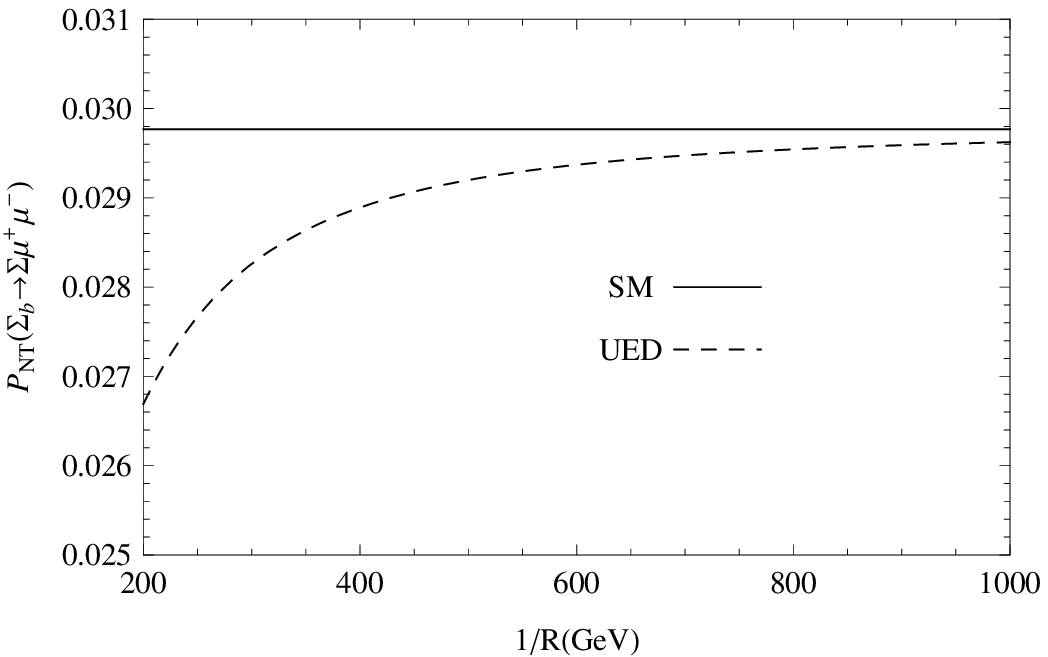,width=0.45\linewidth,clip=} &
\epsfig{file=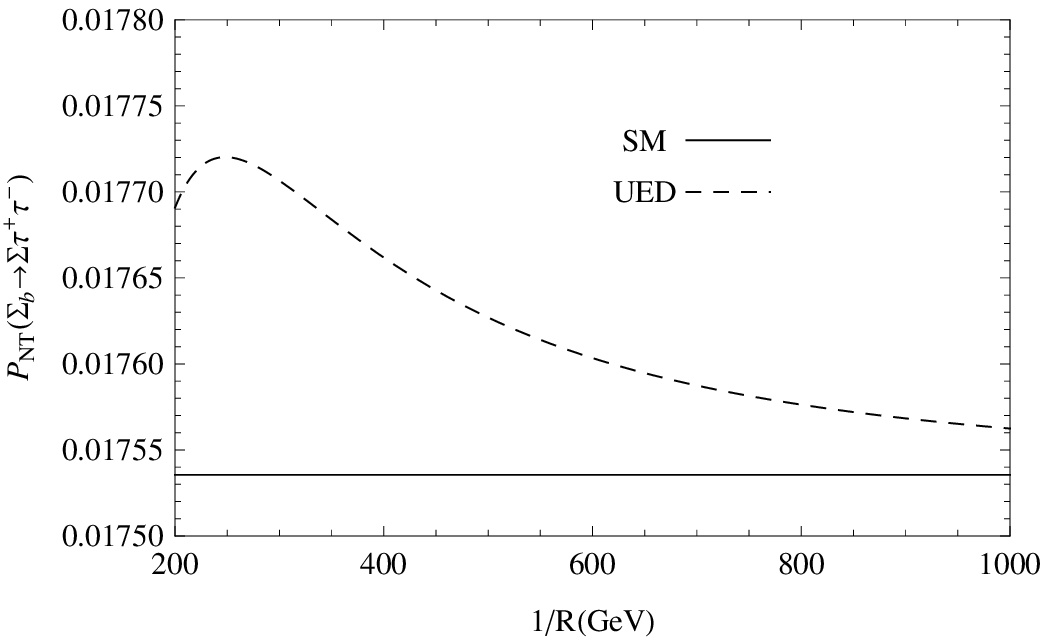,width=0.45\linewidth,clip=}
\end{tabular}
\caption{The $1/R$ dependence of the $P_{NT}(\hat s,1/R)$ at $\hat s=0.5$.}
\end{figure}

\bea \label{PTL} P_{TL}(\hat s, 1/R) \es \frac{16 \pi m_{\Sigma_b}^4
\hat{m}_\ell \sqrt{\lambda} v}{\Delta(\hat s, 1/R) \sqrt{\hat{s}}} \mbox{\rm
Re} \Bigg\{
(1-\hat{r}_\Sigma) \Big( \vel D_1 \ver^2 + \vel E_1 \ver^2 \Big)
+ \hat{s} \Big(A_1 D_1^\ast - B_1 E_1^\ast \Big) \nnb \\
\ar m_{\Sigma_b} \hat{s} \Big[ B_1 D_2^\ast + (A_2 - D_2 + D_3)
E_1^\ast -  A_1 E_2^\ast
- (B_2+E_2-E_3) D_1^\ast \Big] \nnb \\
\ek m_{\Sigma_b}
 \sqrt{\hat{r}_\Sigma} \hat{s}
\Big[ A_1 D_2^\ast + (A_2 - D_2 - D_3) D_1^\ast - B_1 E_2^\ast -
(B_2 + E_2 + E_3) E_1^\ast \Big] \nnb \\
\ek m_{\Sigma_b}^2 \hat{s} (1-\hat{r}_\Sigma) (A_2 D_2^\ast -
B_2 E_2^\ast)
- m_{\Sigma_b}^2 \hat{s}^2 (D_2 D_3^\ast + E_2 E_3^\ast ) \Bigg\},\eea
\begin{figure}[h!]
\centering
\begin{tabular}{ccc}
\epsfig{file=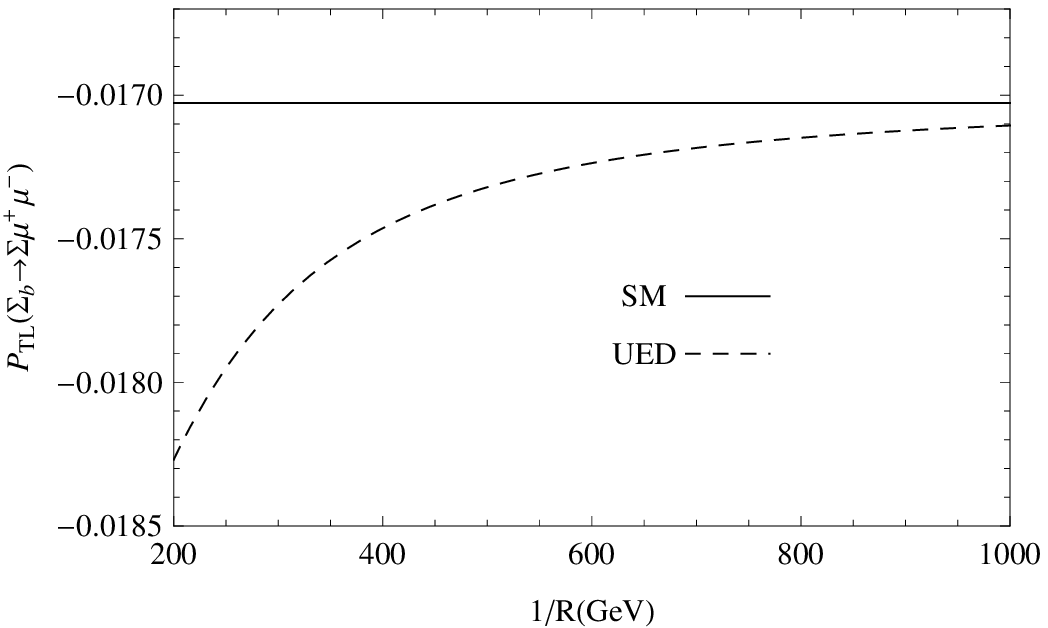,width=0.45\linewidth,clip=} &
\epsfig{file=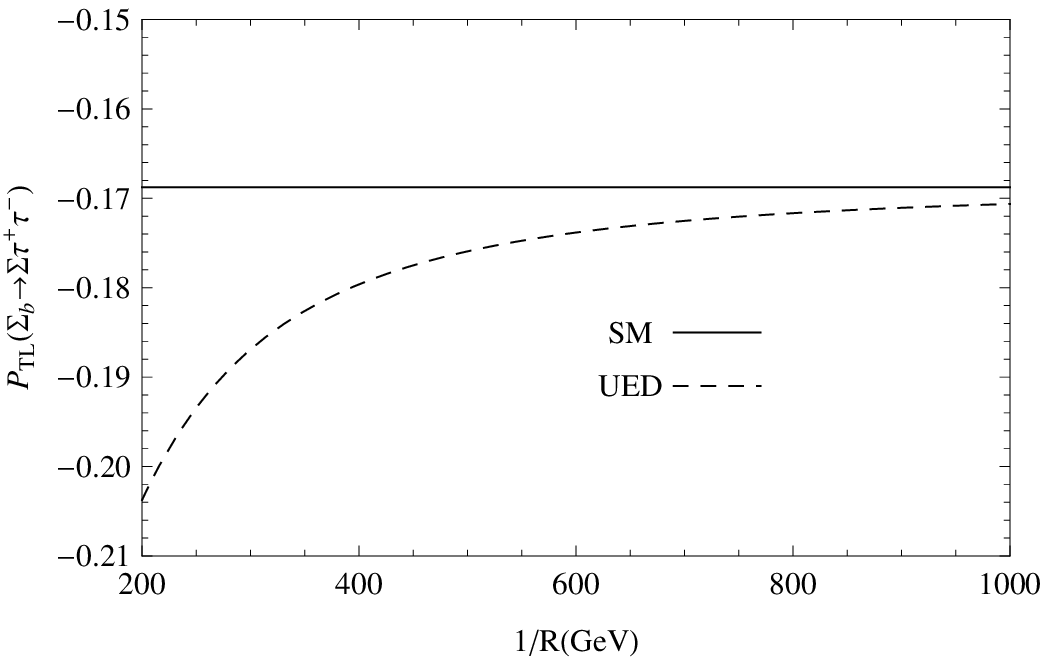,width=0.45\linewidth,clip=}
\end{tabular}
\caption{The $1/R$ dependence of the $P_{TL}(\hat s,1/R)$ at $\hat s=0.5$.}
\end{figure}

\bea \label{PTN} P_{TN}(\hat s, 1/R) \es - \frac{64 m_{\Sigma_b}^4 \lambda
v}{3 \Delta(\hat s, 1/R)} \mbox{\rm Im} \Bigg\{
(A_1 D_1^\ast +B_1 E_1^\ast)
+ m_{\Sigma_b}^2 \hat{s} (A_2^\ast D_2 + B_2^\ast E_2)\Bigg\},
\eea
\begin{figure}[h!]
\centering
\begin{tabular}{ccc}
\epsfig{file=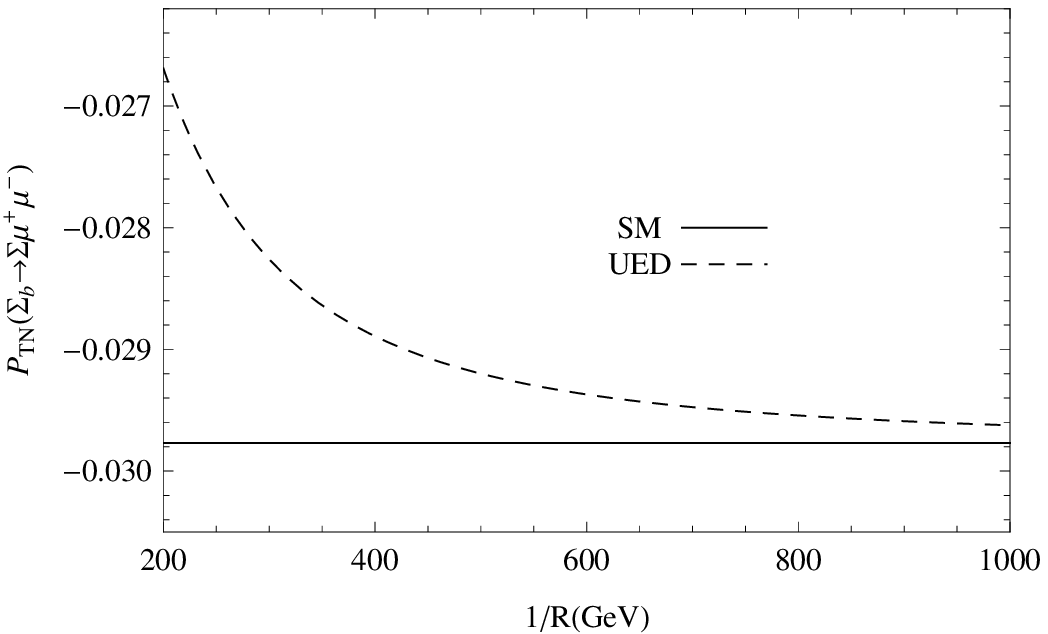,width=0.45\linewidth,clip=} &
\epsfig{file=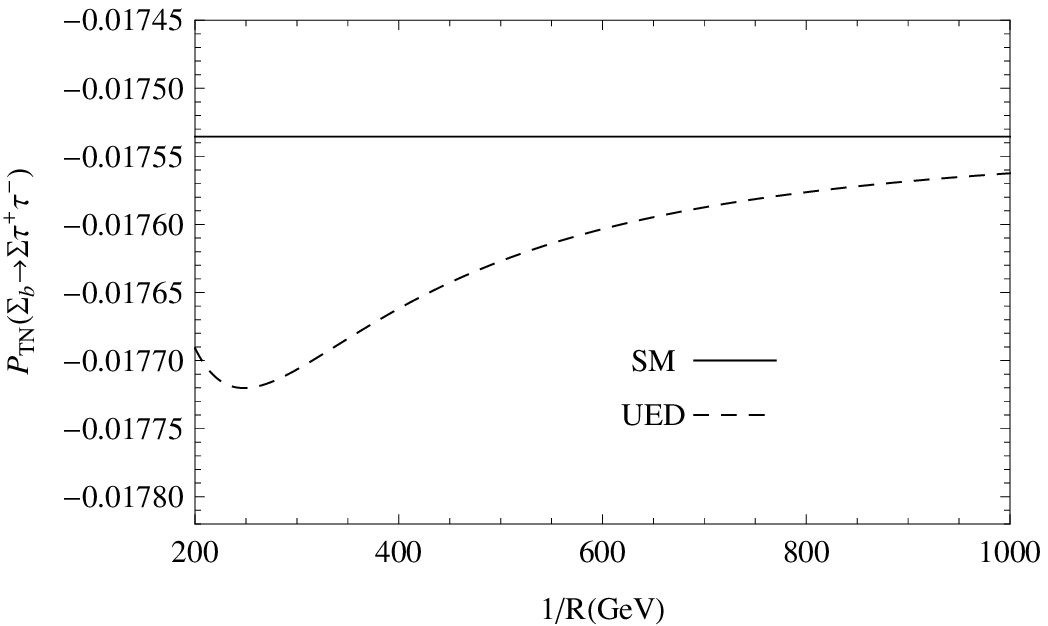,width=0.45\linewidth,clip=}
\end{tabular}
\caption{The $1/R$ dependence of the $P_{TN}(\hat s,1/R)$ at $\hat s=0.5$.}
\end{figure}
\bea \label{PTT} P_{TT}(\hat s, 1/R) \es \frac{32 m_{\Sigma_b}^4}{3 \hat{s}
\Delta(\hat s, 1/R)} \mbox{\rm Re} \Bigg\{
- 24 \hat{m}_\ell^2 \sqrt{r} \hat{s}
( A_1 B_1^\ast + D_1 E_1^\ast ) \nnb \\
\ek 12 m_{\Sigma_b} \hat{m}_\ell^2 \sqrt{r} \hat{s}
(1-r +\hat{s}) (D_1 D_3^\ast + E_1 E_3^\ast)
- 24 m_{\Sigma_b}^2 \hat{m}_\ell^2 \sqrt{r}
\hat{s}^2
( A_2 B_2^\ast + D_3 E_3^\ast ) \nnb \\
\ek 6 m_{\Sigma_b} \hat{m}_\ell^2 \hat{s} \Big[ m_{\Sigma_b}
\hat{s} (1+r-\hat{s}) \Big(\vel D_3 \ver^2 + \vel
E_3 \ver^2 \Big) - 2 \sqrt{r}
(1-r+\hat{s})
(A_1 A_2^\ast + B_1 B_2^\ast)\Big] \nnb \\
\ek 12 m_{\Sigma_b} \hat{m}_\ell^2 \hat{s}
(1-r-\hat{s})
(A_1 B_2^\ast + A_2 B_1 ^\ast + D_1 E_3^\ast + D_3 E_1^\ast) \nnb \\
\ek [ \lambda \hat{s} - 2 \hat{m}_\ell^2 (1 + r^2 -
2 r + r \hat{s} + \hat{s} - 2
\hat{s}^2) ]
\Big( \vel A_1 \ver^2 + \vel B_1 \ver^2 \Big) \nnb \\
\ar m_{\Sigma_b}^2 \hat{s} \{ \lambda \hat{s} + \hat{m}_\ell^2 [4
(1- r)^2 - 2 \hat{s} (1+r) - 2
\hat{s}^2 ]\}
\Big( \vel A_2 \ver^2 + \vel B_2 \ver^2 \Big) \nnb \\
\ar \{ \lambda \hat{s} - 2 \hat{m}_\ell^2 [5 (1-
r)^2 - 7 \hat{s} (1+r) + 2 \hat{s}^2
]\}
\Big( \vel D_1 \ver^2 + \vel E_1 \ver^2 \Big) \nnb \\
\ek m_{\Sigma_b}^2 \lambda \hat{s}^2 v^2 \Big( \vel D_2 \ver^2 +
\vel E_2 \ver^2 \Big) \Bigg\}, \eea
\begin{figure}[h!]
\centering
\begin{tabular}{ccc}
\epsfig{file=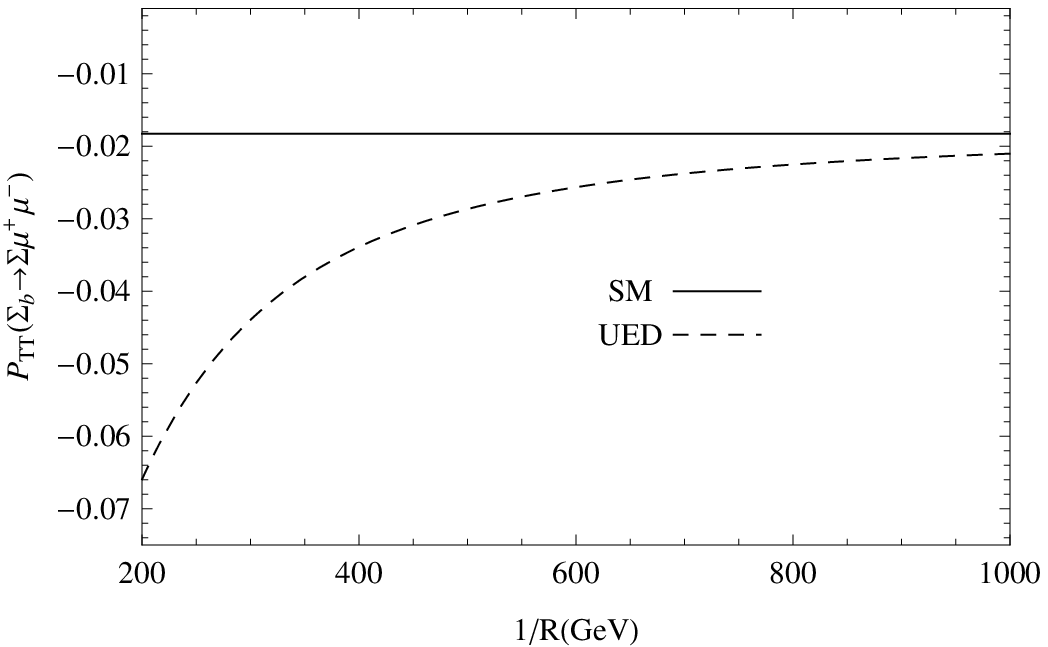,width=0.45\linewidth,clip=} &
\epsfig{file=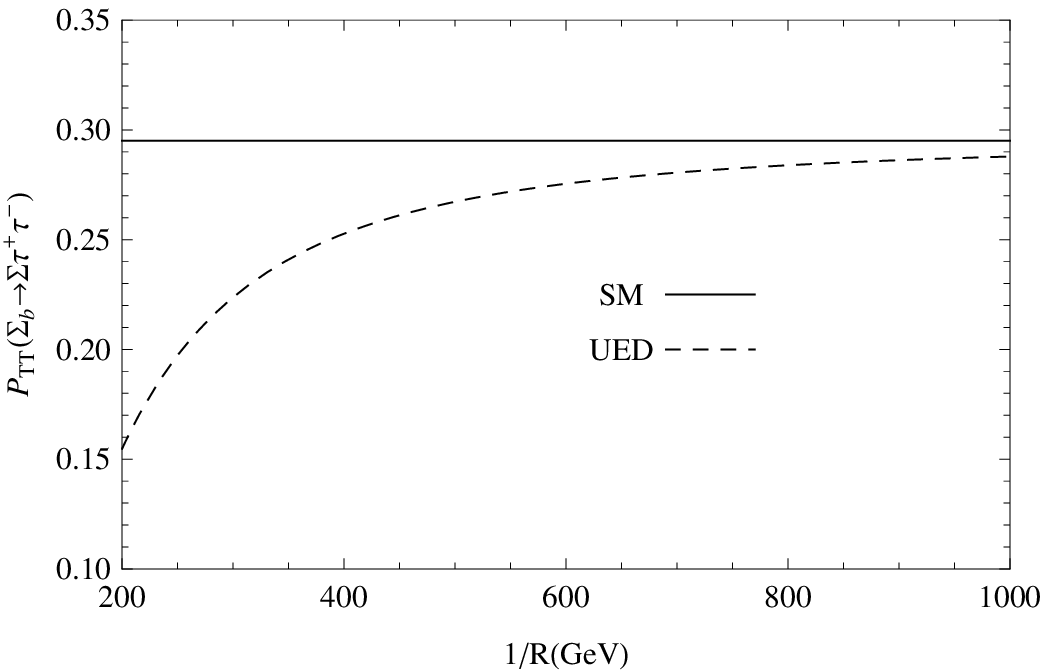,width=0.45\linewidth,clip=}
\end{tabular}
\caption{The $1/R$ dependence of the $P_{TT}(\hat s,1/R)$ at $\hat s=0.5$.}
\end{figure}
where, $\hat m_l=\frac{m_l}{m_{\Sigma_b}}$. The dependence of various double lepton polarization asymmetries are presented in Figures 7-14. Our numerical analysis show that
\begin{itemize}
\item there are also considerable discrepancies between two model predictions at lower values of the compactification scale.
\item The $P_{LL}$, $P_{NN}$, $P_{TT}$ and $P_{LT}$ are very sensitive to new physics effects, while the effects of UED on $P_{TN}$, $P_{TL}$, $P_{NT}$, $P_{NL}$ and $P_{LN}$ are small.
\item Except than the $P_{TT}$, all polarizations have the same sign for both leptons.
\end{itemize}

\subsection{Physical Observables Considering Uncertainties of the Form Factors}
In the previous subsections, we numerically analyzed the physical quantities under consideration and discussed their dependencies on the compactification factor of extra dimension when only the central
values of the form factors are considered. Here,  we discuss how the uncertainties of the form factors  as the main inputs affect the obtained results. For this aim, we present dependencies
of  different physical observables for $\Sigma_{b}\rightarrow \Sigma \ell^{+}\ell^{-}$  on $1/R$ when the errors of the form factors are taken into account in figures 15-27.
\begin{figure}[h!]
\label{error}
\centering
\begin{tabular}{ccc}
\epsfig{file=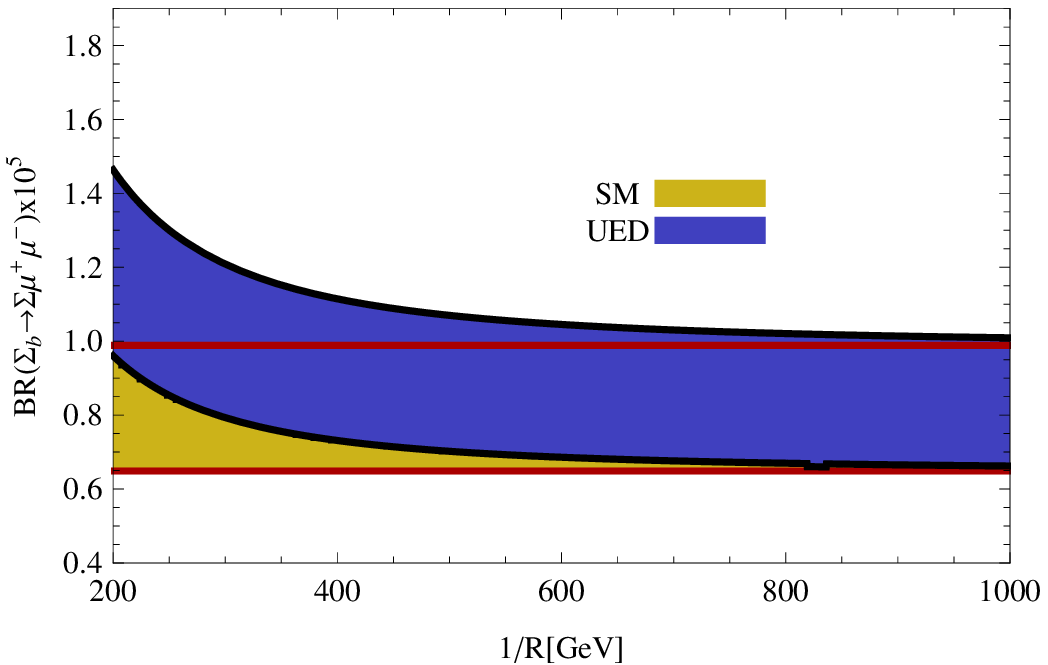,width=0.45\linewidth,clip=} &
\epsfig{file=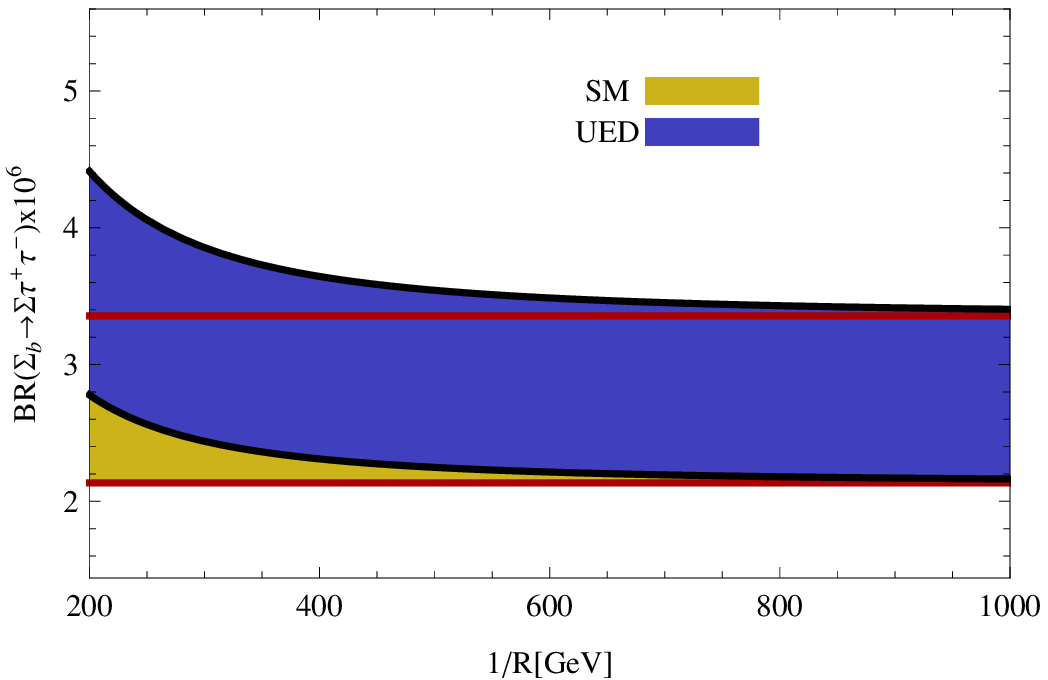,width=0.45\linewidth,clip=}
\end{tabular}
\caption{The $1/R$ dependence of the branching ratio for $\Sigma_{b}\rightarrow \Sigma \ell^{+}\ell^{-}$ at $\hat s=0.5$ when errors of the form factors are taken into account. The Brown-Yellow bands surrounded by red lines refer to the SM,
 while the blue bands surrounded by black lines denote the UED results.}
\end{figure}
\begin{figure}[h!]
\centering
\begin{tabular}{ccc}
\epsfig{file=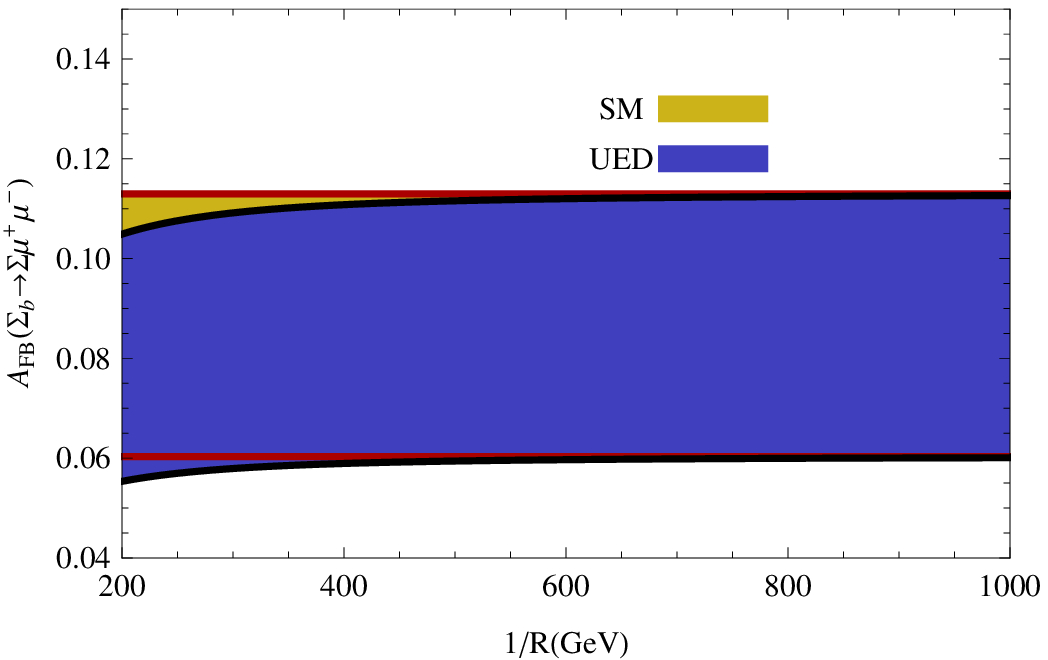,width=0.45\linewidth,clip=} &
\epsfig{file=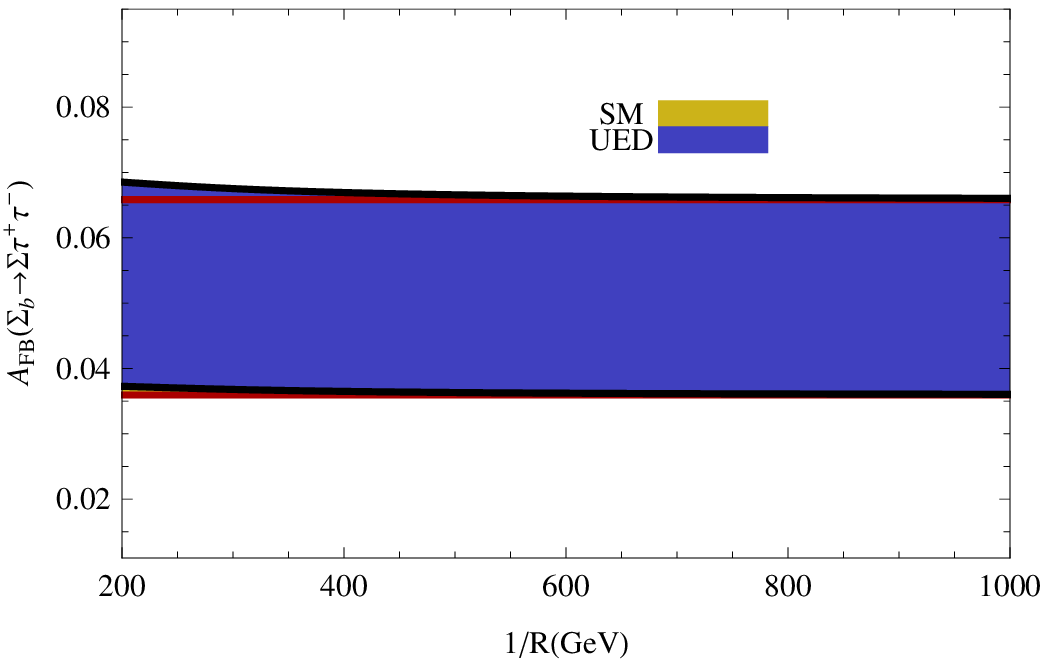,width=0.45\linewidth,clip=}
\end{tabular}
\caption{The same as figure 15 but for  forward-backward asymmetry.}
\end{figure}
\begin{figure}[h!]
\centering
\begin{tabular}{ccc}
\epsfig{file=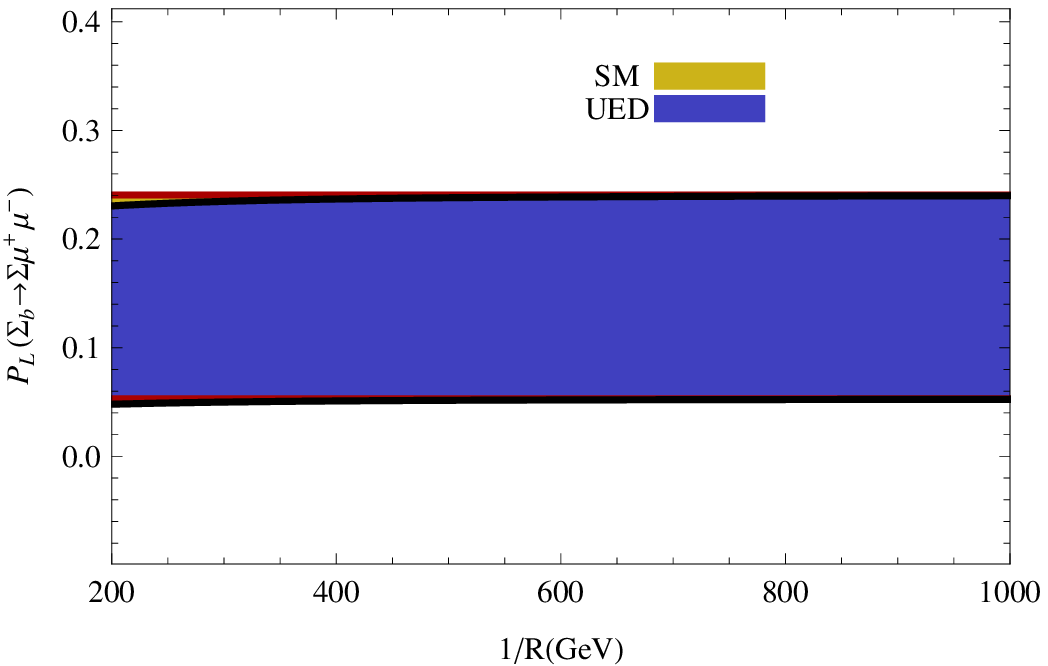,width=0.45\linewidth,clip=} &
\epsfig{file=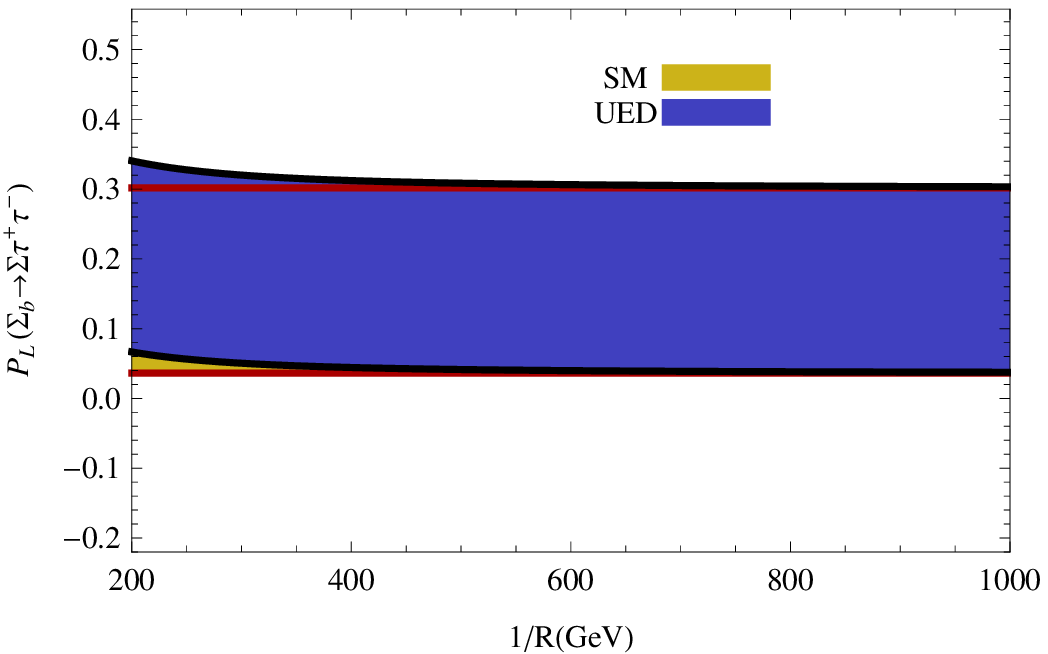,width=0.45\linewidth,clip=}
\end{tabular}
\caption{The same as figure 15 but for  $P_L$.}
\end{figure}
\begin{figure}[h!]
\centering
\begin{tabular}{ccc}
\epsfig{file=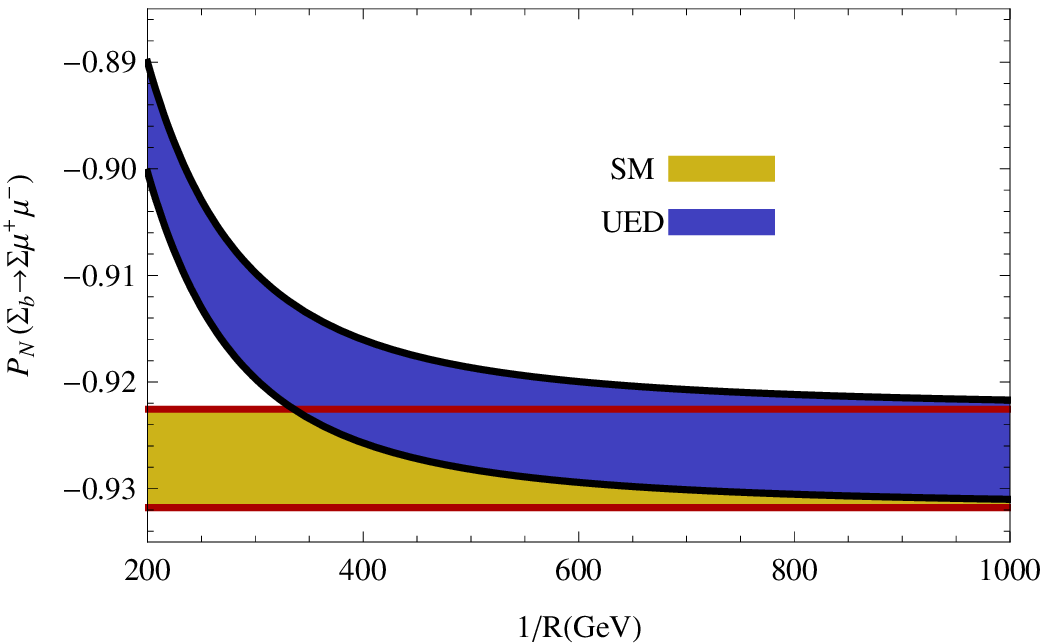,width=0.45\linewidth,clip=} &
\epsfig{file=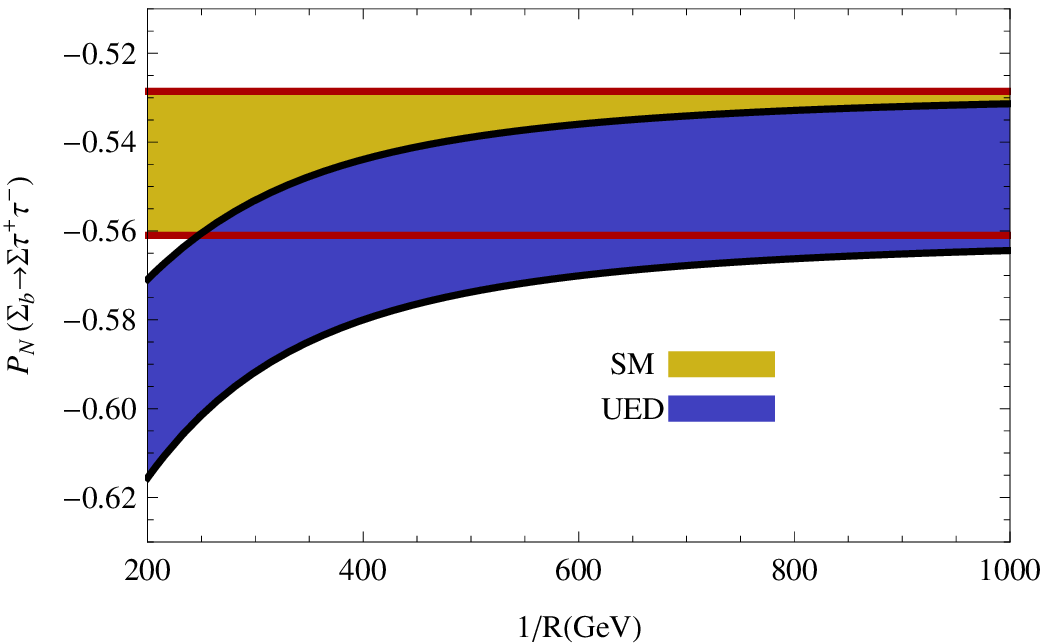,width=0.45\linewidth,clip=}
\end{tabular}
\caption{The same as figure 15 but for  $P_N$.}
\end{figure}
\begin{figure}[h!]
\centering
\begin{tabular}{ccc}
\epsfig{file=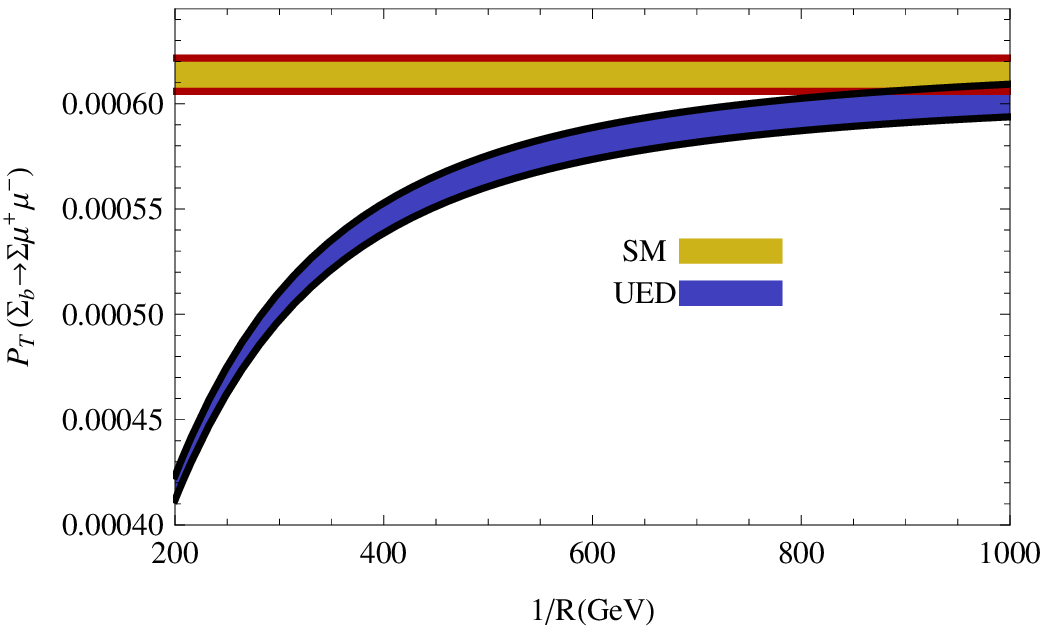,width=0.45\linewidth,clip=} &
\epsfig{file=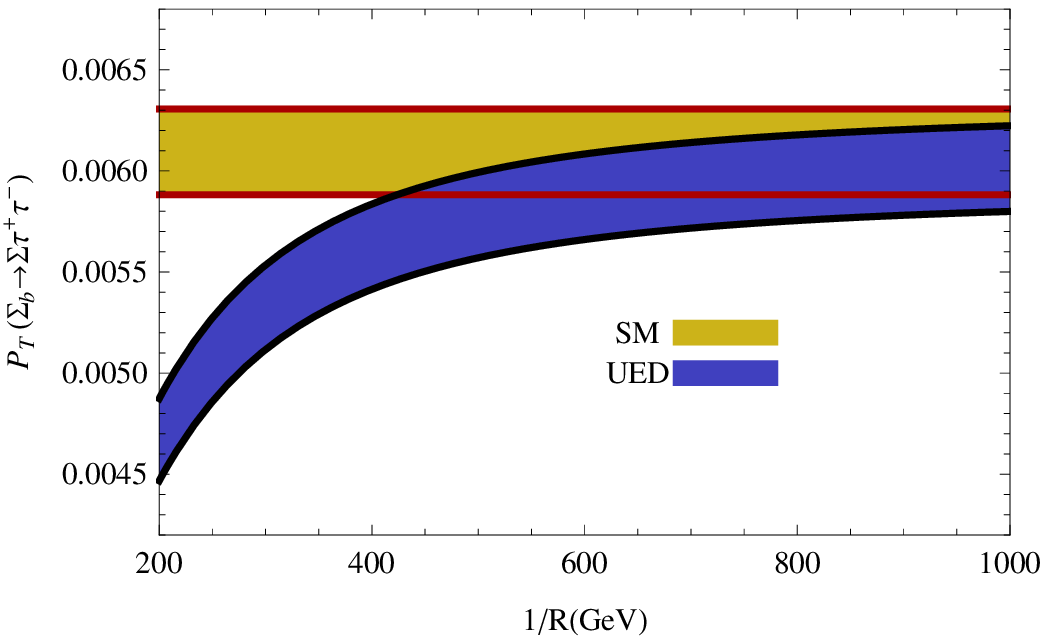,width=0.45\linewidth,clip=}
\end{tabular}
\caption{The same as figure 15 but for  $P_T$.}
\end{figure}
\begin{figure}[h!]
\centering
\begin{tabular}{ccc}
\epsfig{file=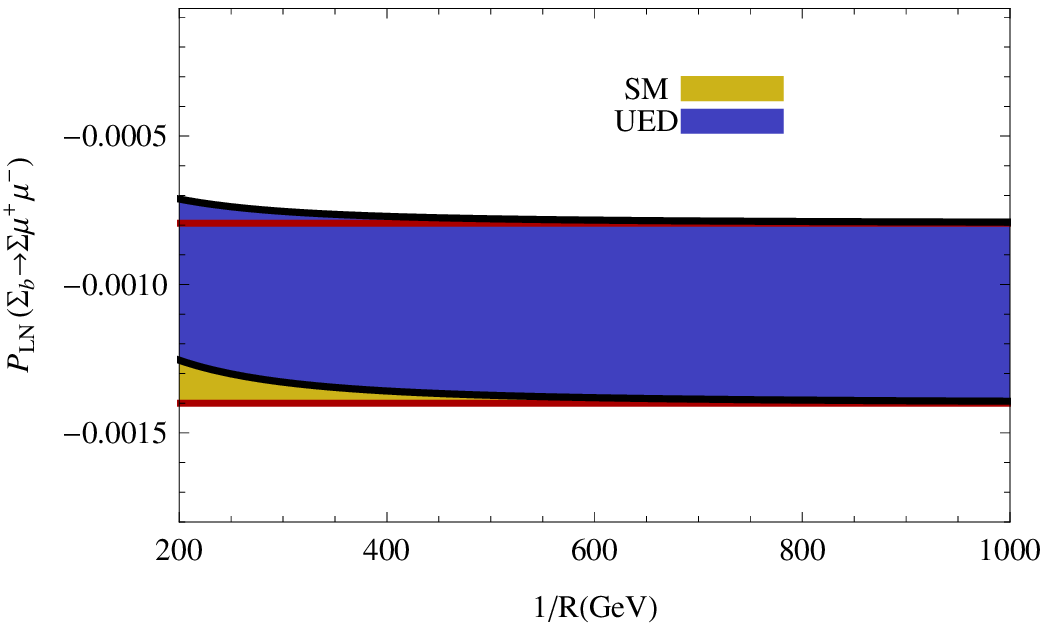,width=0.45\linewidth,clip=} &
\epsfig{file=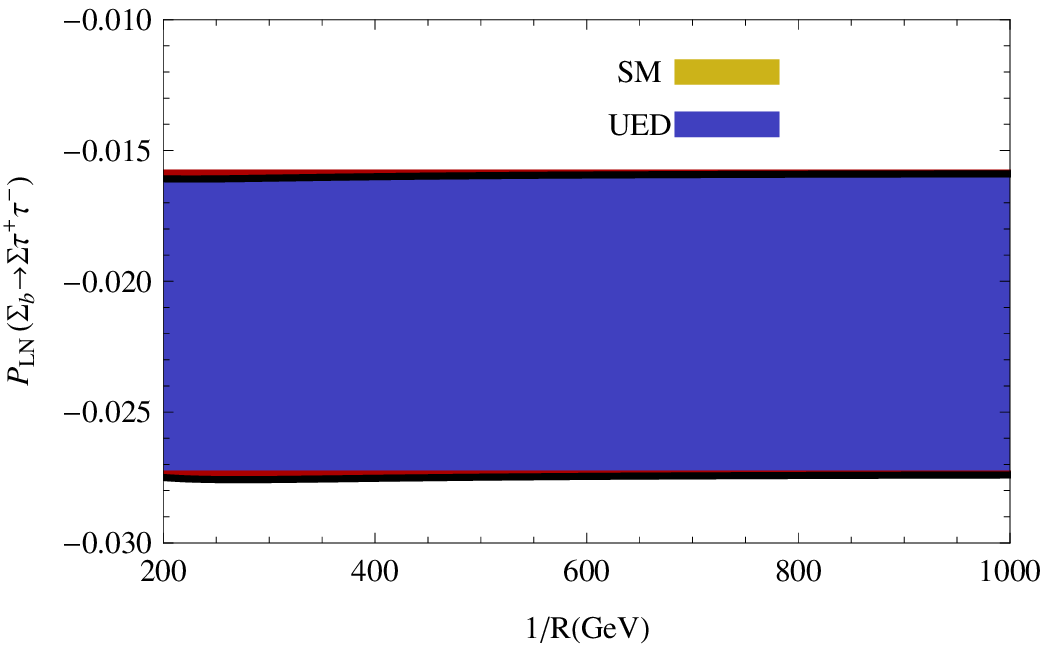,width=0.45\linewidth,clip=}
\end{tabular}
\caption{The same as figure 15 but for $P_{LN}$.}
\end{figure}
\begin{figure}[h!]
\centering
\begin{tabular}{ccc}
\epsfig{file=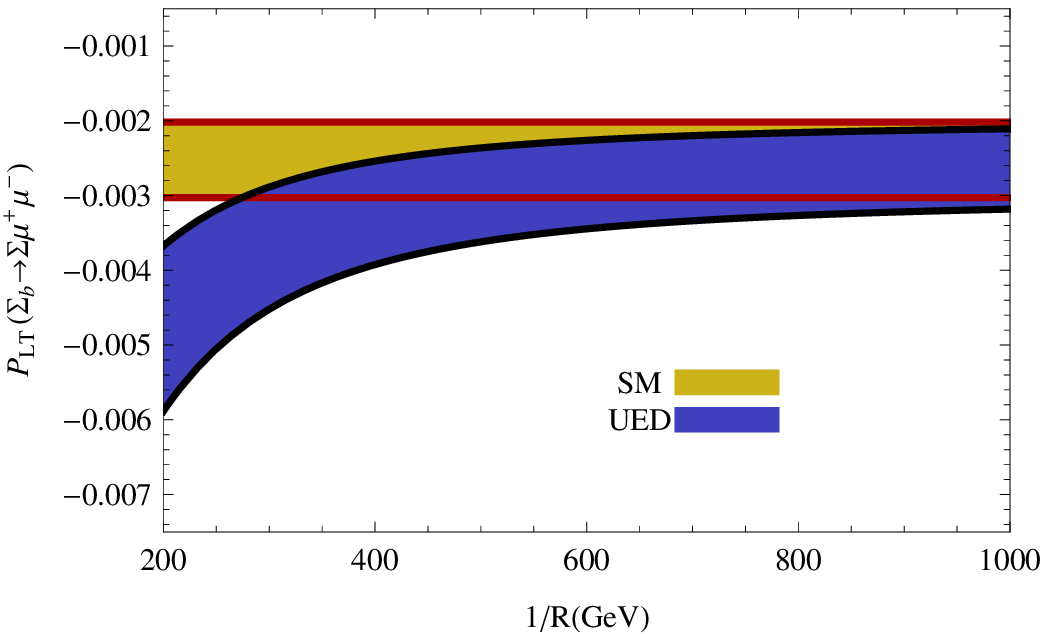,width=0.45\linewidth,clip=} &
\epsfig{file=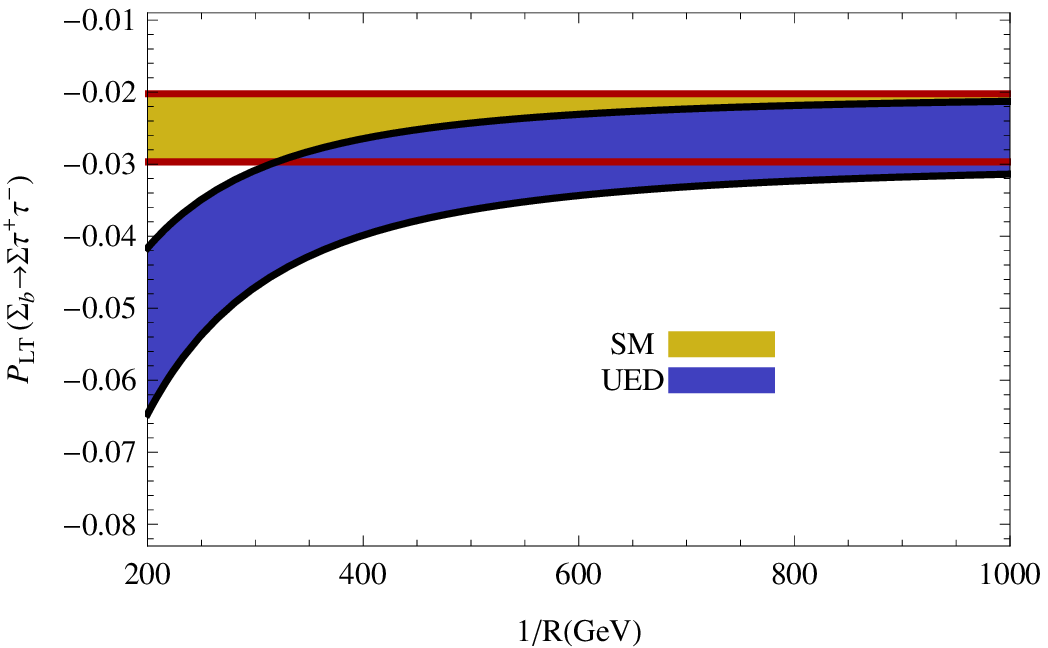,width=0.45\linewidth,clip=}
\end{tabular}
\caption{The same as figure 15 but for $P_{LT}$.}
\end{figure}
\begin{figure}[h!]
\centering
\begin{tabular}{ccc}
\epsfig{file=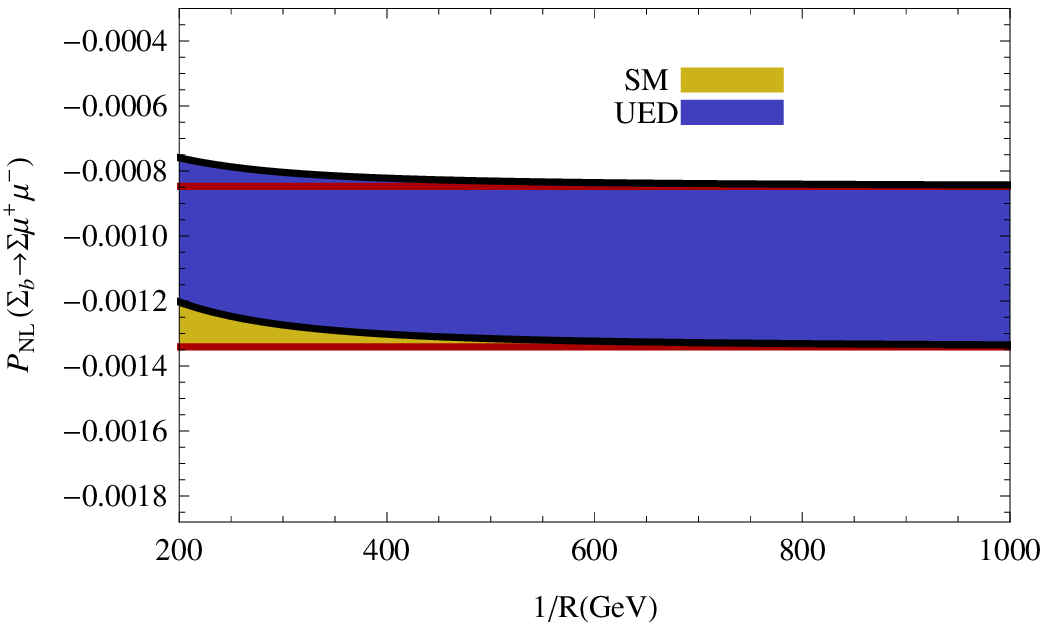,width=0.45\linewidth,clip=} &
\epsfig{file=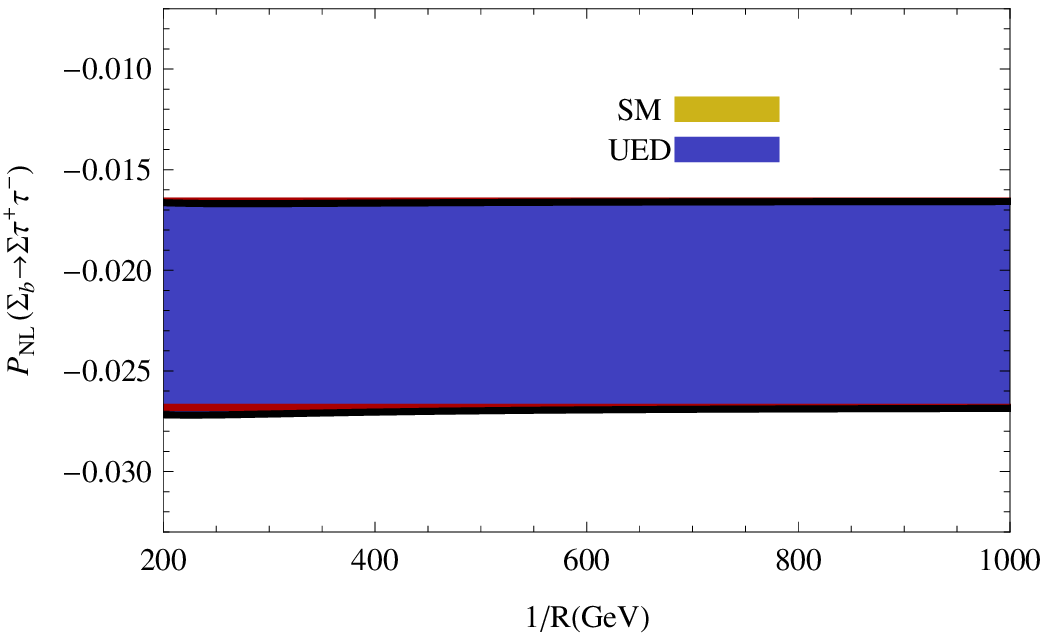,width=0.45\linewidth,clip=}
\end{tabular}
\caption{The same as figure 15 but for $P_{NL}$.}
\end{figure}
\begin{figure}[h!]
\centering
\begin{tabular}{ccc}
\epsfig{file=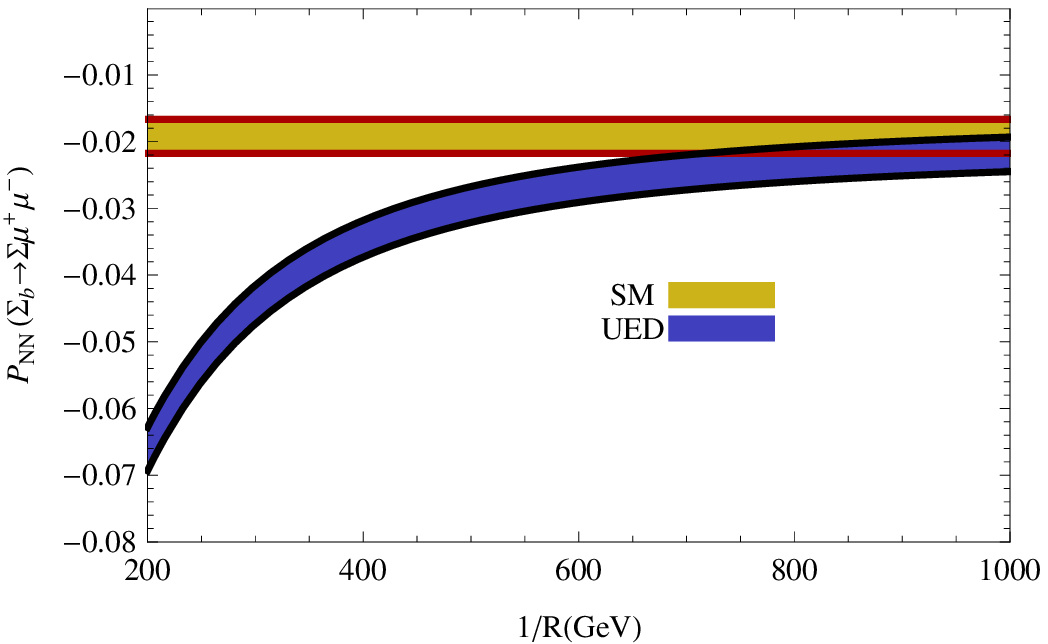,width=0.45\linewidth,clip=} &
\epsfig{file=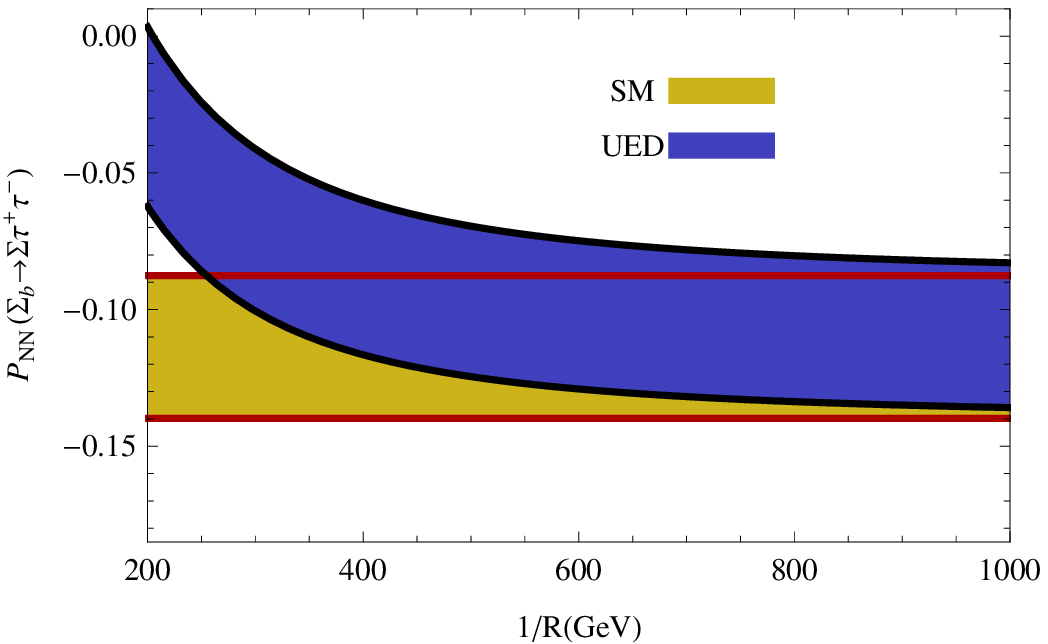,width=0.45\linewidth,clip=}
\end{tabular}
\caption{The same as figure 15 but for $P_{NN}$.}
\end{figure}
\begin{figure}[h!]
\centering
\begin{tabular}{ccc}
\epsfig{file=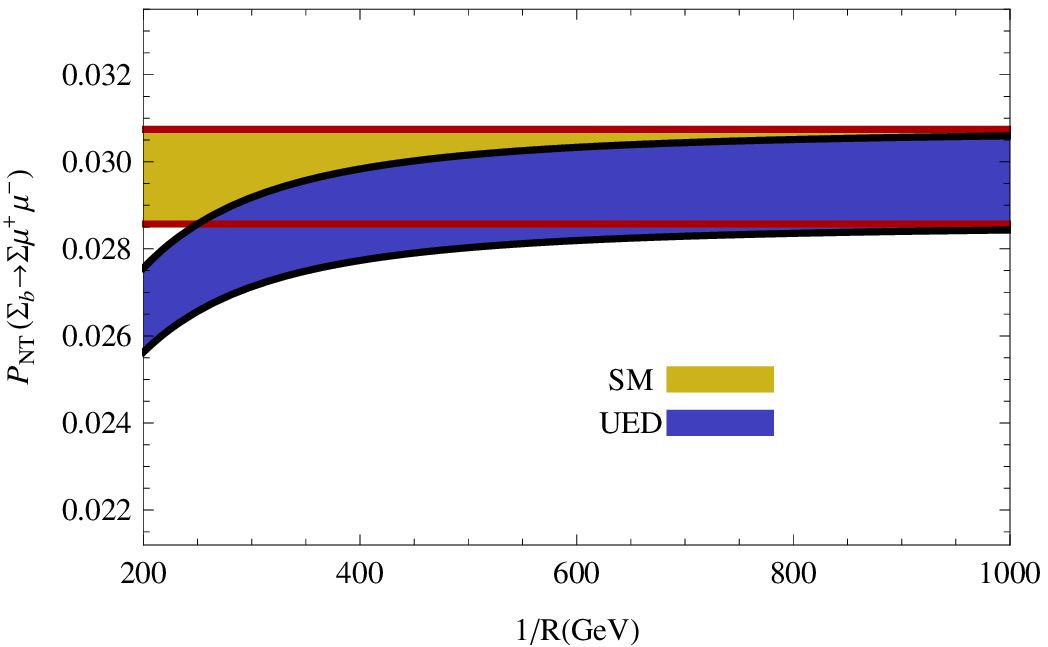,width=0.45\linewidth,clip=} &
\epsfig{file=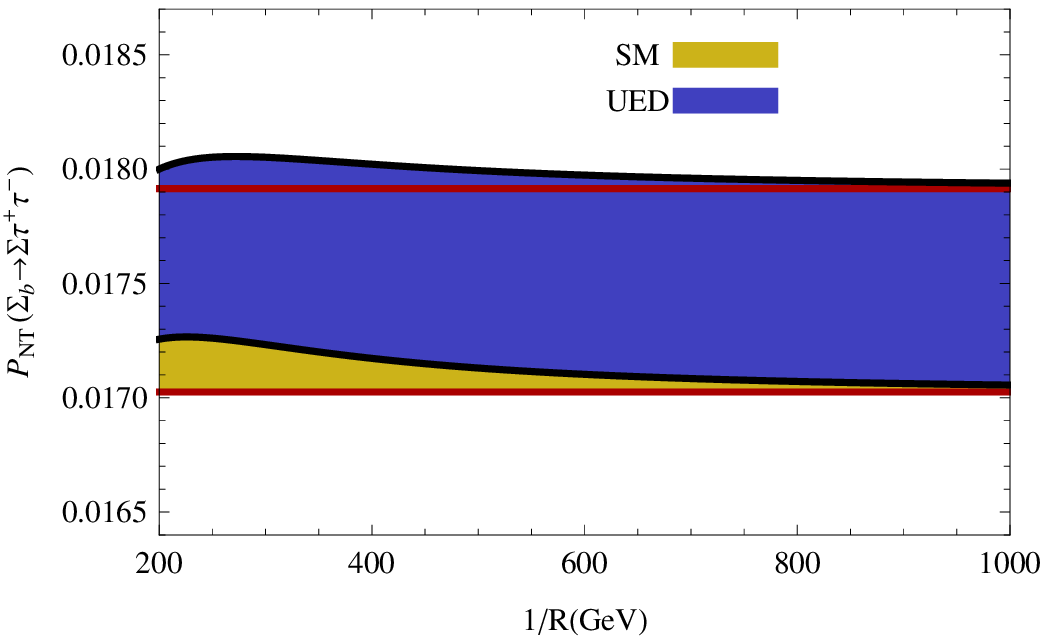,width=0.45\linewidth,clip=}
\end{tabular}
\caption{ The same as figure 15 but for $P_{NT}$.}
\end{figure}
\begin{figure}[h!]
\centering
\begin{tabular}{ccc}
\epsfig{file=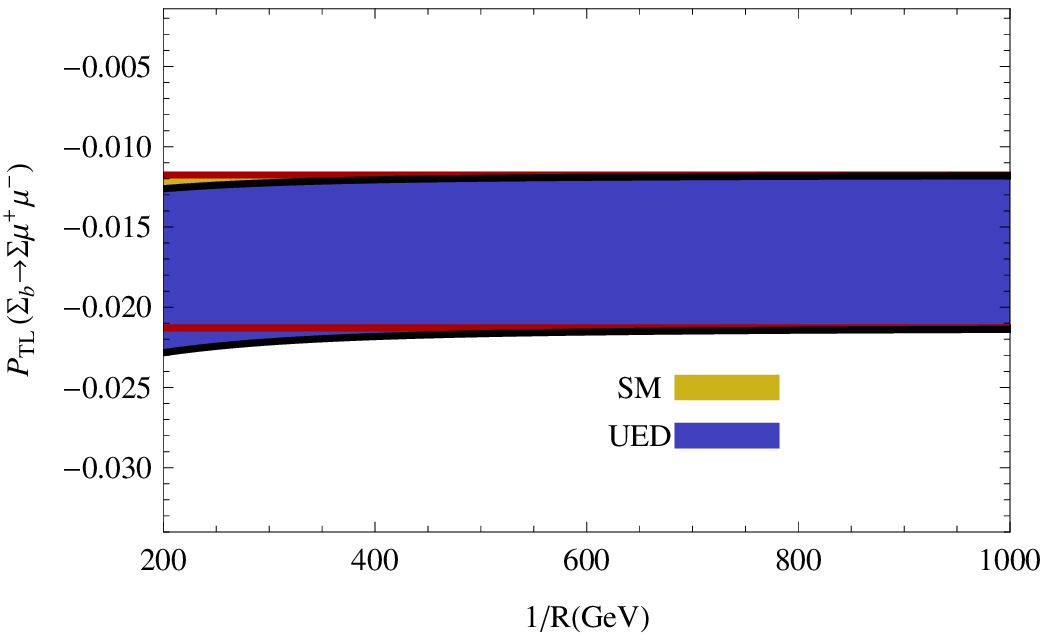,width=0.45\linewidth,clip=} &
\epsfig{file=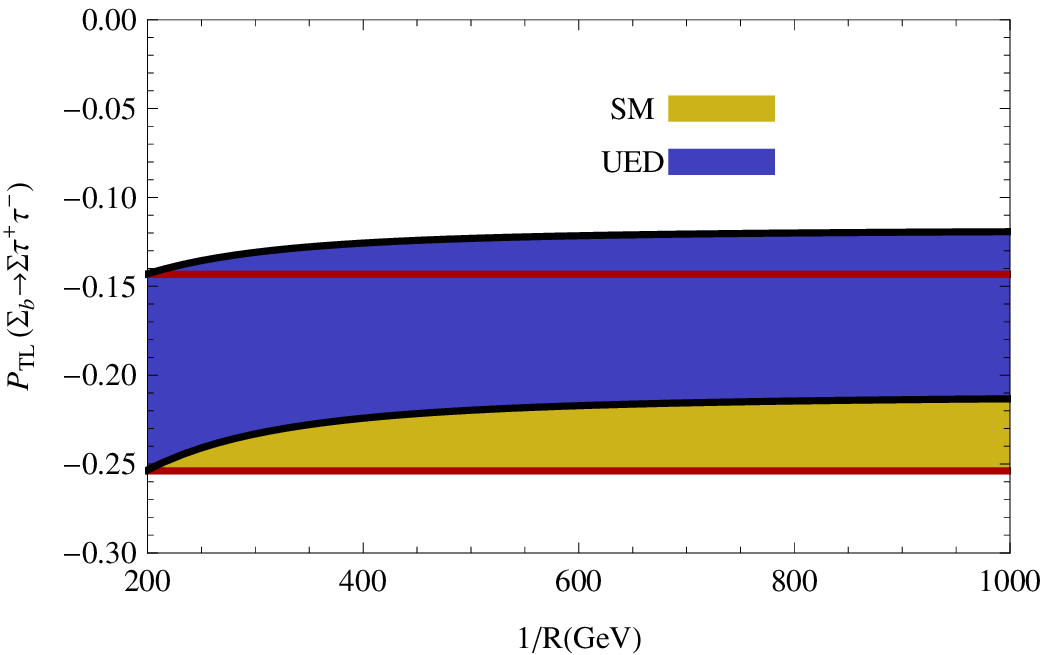,width=0.45\linewidth,clip=}
\end{tabular}
\caption{The same as figure 15 but for $P_{TL}$.}
\end{figure}
\begin{figure}[h!]
\centering
\begin{tabular}{ccc}
\epsfig{file=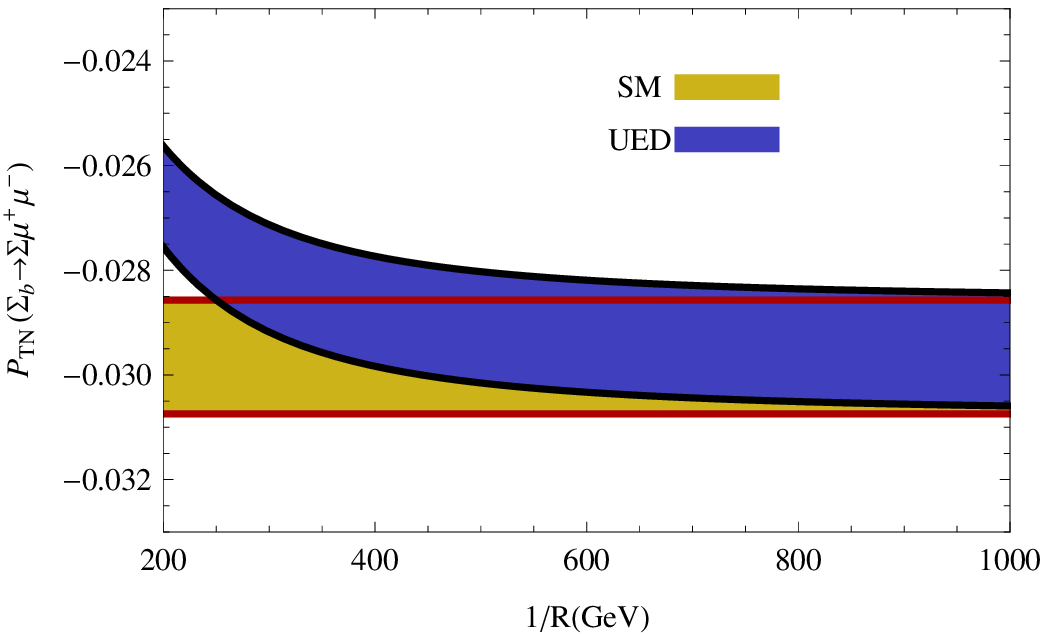,width=0.45\linewidth,clip=} &
\epsfig{file=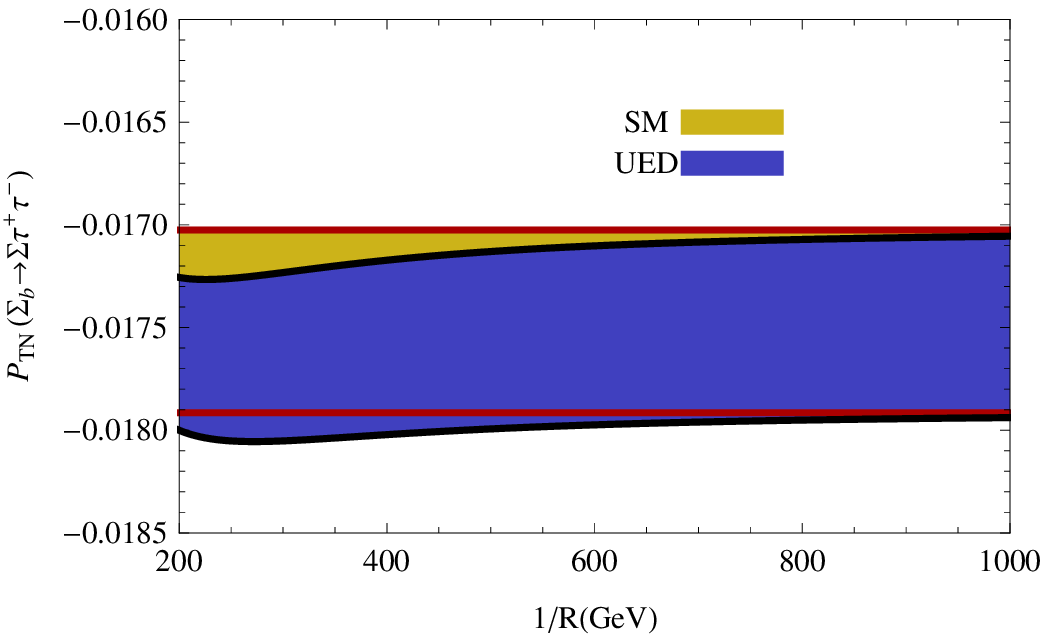,width=0.45\linewidth,clip=}
\end{tabular}
\caption{The same as figure 15 but for $P_{TN}$.}
\end{figure}
\begin{figure}[h!]
\centering
\begin{tabular}{ccc}
\epsfig{file=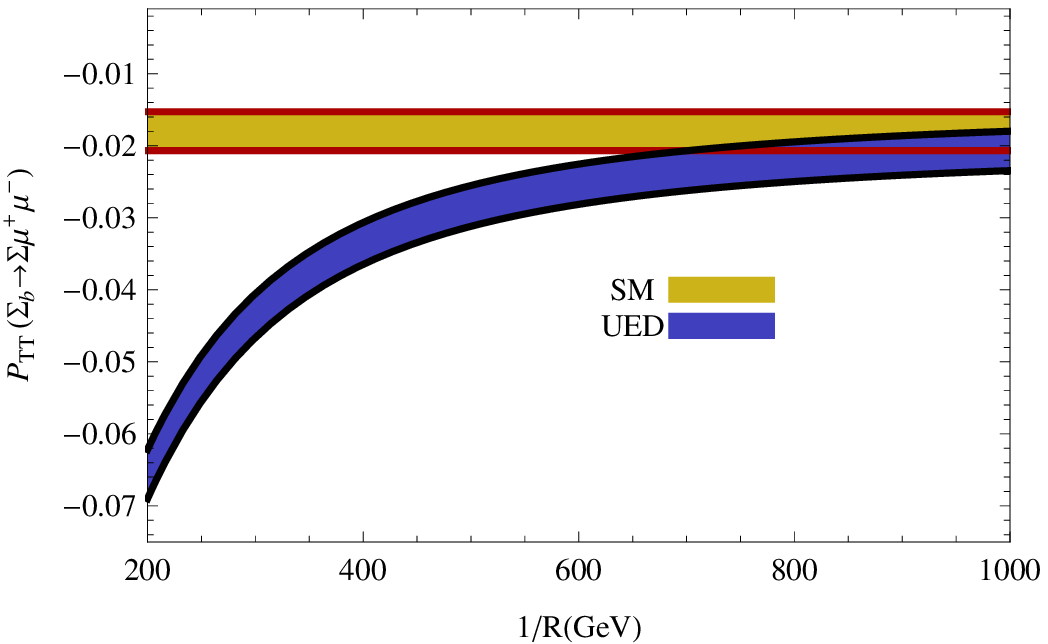,width=0.45\linewidth,clip=} &
\epsfig{file=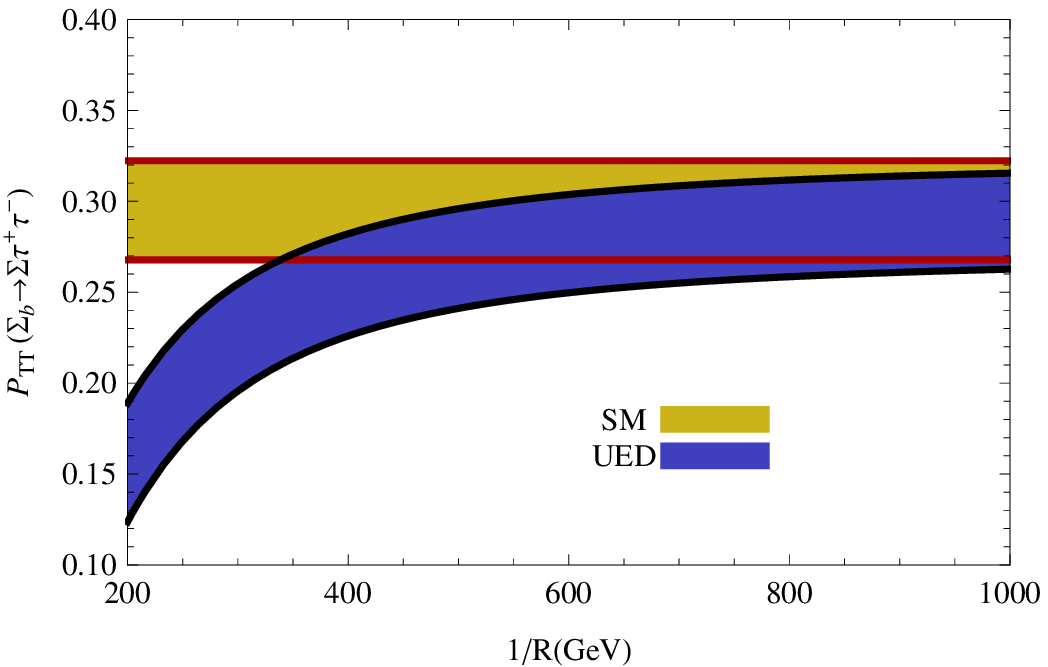,width=0.45\linewidth,clip=}
\end{tabular}
\caption{The same as figure 15 but for $P_{TT}$.}
\end{figure}

From these figures we see that in all cases, the SM and UED bands intersect each other in some regions. In some cases like $P_{N}$  at $\mu$ channel as well as $P_{T}$, $P_{TT}$, $P_{NN}$ and $P_{LT}$ at both 
lepton channels, the errors of the form factors can not kill the differences between the UED and SM predictions at small values of the compactification factor. In the case of forward-backward asymmetry and longitudinal 
baryon polarization for both leptons;  $P_{LN}$ and $P_{NL}$ at $\tau$ channel as well as $P_{TL}$ at $\mu$ channel, the differences between two model predictions are killed by the uncertainties of the form factors.
For the other cases like branching ratio at both lepton channels; 
$P_{LN}$ and $P_{NL}$ at $\mu$ channel; and $P_{NT}$,  $P_{TN}$ and $P_{TL}$ at $\tau$ channel, we have intermediate situation and  see some small but considerable regions out of the intersection parts of the SM and 
UED predictions.

\section{Conclusion}
In the present study, we found a lower limit for the compactification scale of extra dimension comparing the recent experimental data on the branching ratio of baryonic FCNC 
$\Lambda _b\rightarrow \Lambda \mu^+ \mu^-$ transition and our previous theoretical work. We put an approximately $250~GeV$ for lower limit of the compactification factor in the presence of a single UED.
 This limit is in a 
good consistency with the lower limit very recently obtained via comparison between the experimental data and theoretical results (containing a single UED) on the branching fraction of
 the mesonic $ B \rightarrow K \eta \gamma$ \cite{colangeloR} channel. Our result  is also  comparable with some other limits previously obtained in other mesonic channels 
 as well as some electroweak 
precision tests \cite{ikiuc,ACD,Appelquist}. 
However, our lower limit on $1/R$ is small compared to the one also obtained in  $B \rightarrow K \eta \gamma$ channel but in the presence of 2 UEDs as well as obtained from some other mesonic decay channels,
 electroweak precision tests,
some cosmological constraints and ATLAS results discussed in section 2 \cite{Gogoladze:2006br,Cembranos:2006gt,Haisch:2007vb,ATLAS}. To improve our limit,   we need the expressions of the Wilson coefficients $C_9^{eff}$ and  $C_{10}$ calculated in the
presence of 2 UEDs, theoretically. From the experimental point of view, we are waiting for the results of LHCb on the physical observables related to the 
 $\Lambda _b\rightarrow \Lambda \mu^+ \mu^-$ channel to confirm the CDF data \cite{CDF}.

In the second part, we have analyzed the other baryonic  $\Sigma_b \rightarrow \Sigma
\ell^+ \ell^-$ decay channel also in UED scenario. Using the form factors recently available and calculated via light cone QCD sum rules in full theory, we have discussed  sensitivity of  many related  physical
observables such as  branching ratio, forward-backward asymmetry, baryon polarizations and double lepton polarization asymmetries on the compactification factor of extra dimension. We have observed over all
sizable discrepancies between  the UED and SM predictions at lower values of the compactification scale when we considered the central values of the form factors as the main inputs. Although these discrepancies
are killed by uncertainties of the form factors for some cases discussed in the body text, for many observables we have still considerable differences between two model predictions. These can be considered as indications for existing the KK modes and extra dimensions should we search for them
at hadron colliders. The order of branching fraction in $\Sigma_b \rightarrow \Sigma
\ell^+ \ell^-$ decay channel indicates that this channel is accessible at  LHC.

\section{Acknowledgement}
We thank V. N. \c Seno\u guz for useful discussions on lower limit of compactification scale.

\end{document}